\documentclass[journal]{IEEEtran}
\usepackage{cite}
\ifCLASSINFOpdf
  \usepackage[pdftex]{graphicx}
  \graphicspath{{../images/}}
  \DeclareGraphicsExtensions{.pdf,.jpeg,.png}
\else
  \usepackage[dvips]{graphicx}
  \graphicspath{{../images/}}
  \DeclareGraphicsExtensions{.eps}
\fi
\usepackage{comment}

\usepackage{booktabs}

\usepackage{tabularx}

\usepackage{multirow}

\usepackage{pgfplots}

\usepackage{caption}

\usepackage{subcaption}

\usepackage{adjustbox}

\usepackage{booktabs}

\usepackage{geometry}
\usepackage{float}
\usepackage{stfloats}
\usepackage{amsmath,mathabx}
\usepackage{breqn}
\usepackage{colortbl}
\usepackage{xcolor}

\usepackage{array}

\usepackage{pdfpages}

\usepackage{lipsum} % for placeholder text

\definecolor{lightblue}{rgb}{0.9,0.9,1}

\begin{document}
\title{Preictal Period Optimization for Deep Learning-Based Epileptic Seizure Prediction}

\author{Petros~Koutsouvelis,
        Bartlomiej~Chybowski,
        Alfredo~Gonzalez-Sulser,
        Shima~Abdullateef,
        Javier~Escudero
    
\thanks{P.K.~is with the Department of Radiation Oncology (Maastro), GROW Research Institute for Oncology and Reproduction, Maastricht University Medical Centre+, Maastricht, The Netherlands}
\thanks{B.C., A.G.S., and J.E.~are with the Muir Maxwell Epilepsy Centre, University of Edinburgh, UK.}
\thanks{S.A. is with the Centre for Medical Informatics, Usher Institute, Medical School, University of Edinburgh, UK.}
\thanks{B.C.~and J.E.~are with the Institute for Imaging, Data and Communications, School of Engineering, University of Edinburgh, UK.}}% <-this % stops a space

\maketitle

%Last sentence is an example, can be removed/ changed. 
\begin{abstract}
Accurate prediction of epileptic seizures could prove critical for improving patient safety and quality of life in drug-resistant epilepsy. Although deep learning-based approaches have shown promising seizure prediction performance using scalp electroencephalogram (EEG) signals, substantial limitations still impede their clinical adoption. Furthermore,
identifying the optimal preictal period (OPP) for labeling EEG segments remains a challenge. Here, we not only develop a competitive deep learning model for seizure prediction but, more importantly, leverage it to demonstrate a methodology to comprehensively evaluate the predictive performance in the seizure prediction task. For this, we introduce a CNN-Transformer deep learning model to detect preictal spatiotemporal dynamics, alongside a novel Continuous Input-Output Performance Ratio (CIOPR) metric to determine the OPP. We trained and evaluated our model on 19 pediatric patients of the open-access CHB-MIT dataset in a subject-specific manner. Using the OPP of each patient, preictal and interictal segments were correctly identified with an average sensitivity of 99.31\%, specificity of 95.34\%, AUC of 99.35\%, and F1-score of 97.46\%, while prediction time averaged 76.8 minutes before onset. Notably, our novel CIOPR metric allowed outlining the impact of different preictal period definitions on prediction time, accuracy, output stability, and transition time between interictal and preictal states in a comprehensive and quantitative way and highlighted the importance of considering both inter- and intra-patient variability in seizure prediction.
\end{abstract}


% \fi
\IEEEpeerreviewmaketitle

\section{Introduction}
\IEEEPARstart{E}{pilepsy} stands as one of the most prevalent neurological conditions worldwide, impacting an estimated 65 million individuals. This disorder is defined by recurrent epileptic seizures — unprovoked surges of electrical activity in the brain, leading to transient alterations in behavior or consciousness. Patients span diverse socio-demographic backgrounds, with a lifetime prevalence rate of $3\%$. Notably, an approximate of $ 30\%$ of patients suffer from drug-resistant epilepsy \cite{milligan2021epilepsy}.

An inherent risk factor contributing to mortality among epileptic patients is the unpredictable nature of seizure occurrences, leading to elevated levels of uncertainty, anxiety, social stigma, distress, and potentially hazardous situations among patients \cite{malik2022perceived}. These challenges not only diminish their quality of life but also underscore the critical need for effective seizure management strategies. 

Improvements in predicting epileptic seizures would significantly reduce the burden of uncertainty for patients, enabling them, along with caregivers and clinicians, to take preparatory actions, prevent injuries, and perform timely interventions \cite{wristband_sensor,DLEpilepsy}. Epileptic seizures could be predicted by identifying the preictal state, a period preceding seizure onset (ictal state) characterized by distinct morphological EEG differences from the interictal (normal) state. It can typically last from several minutes to a few hours, varying upon seizures and individuals \cite{OPP}.

\subsection{Deep learning for epileptic seizure prediction}
\label{sub:introDL}
Deep learning algorithms have gained increasing popularity in epileptic seizure prediction owing to their ability to learn complex features from data \cite{DLEpilepsy}. Convolutional Neural Networks (CNN) have demonstrated particular efficacy in automatically extracting underlying patterns of preictal activity from raw EEG signals \cite{Daoud}, EEG wavelets \cite{Khan}, time-frequency data matrices \cite{timefrequency}, and connectivity-based measures \cite{Wang}. Recurrent Neural Networks (RNN), including Gated Recurrent Units (GRU), Long Short-Term Memory (LSTM), and Bi-Directional LSTM (BiLSTM), have been successfully employed to capture temporal dependencies in preictal EEG feature vectors \cite{LSTM, LSTM2}. However, when applied to raw EEG input, RNNs exhibit inferior performance compared to CNNs \cite{CNNAttentionGRU}. Various studies \cite{Daoud, CNNAttentionGRU, CNNGRUAttention} have proposed hybrid models combining the advantages of both architectures with superior overall performance. While direct comparison of reported performances across different studies is not straightforward due to differences in experimental setups, ablation studies within a given work can showcase performance improvements due to specific design choices. Authors in \cite{Daoud} demonstrated that adding a Bi-LSTM layer to a CNN model can improve the accuracy of a subject-specific classifier from 94.1\% to 99.7\%. Similarly, ablation analysis in a cross-patient setting \cite{CNNGRUAttention} showed that adding CNN layers prior to a  GRU classifier led to a 4\% increase in accuracy, illustrating the superiority of hybrid models. 

Attention mechanisms, especially within transformer models, have improved the ability to learn data relationships irrespective of sequence distance, a significant limitation in traditional LSTM applications \cite{attention, attentionAdvantages}. Integrating CNNs with transformers has led to robust models that effectively predict seizures from both raw EEG  \cite{rawCNNTransformer, rawCNNTransformer2} and EEG connectivity maps \cite{CMT}, sometimes combined with additional GRU layers \cite{CNNGRUAttention, CNNAttentionGRU}. Adding a self-attention layer to a CNN-GRU architecture boosted cross-patient classification accuracy from 75.6\% to 82.9\% \cite{CNNGRUAttention}.

\subsection{Determining the preictal period length}
\label{sub:introPreictal}
Despite these technological advances, accurately identifying the optimal preictal period (OPP) \cite{OPP} for labeling EEG segments remains a formidable challenge. This difficulty stems from the gradual nature of the transition into the preictal state and the significant intra- and inter-patient variability in ictogenesis \cite{preictalTransition, preictalConnectivity}. Approaches to determining the OPP are typically categorized into pre-training and post-training. The former involves defining the preictal length for each seizure before model training through the analysis of EEG markers. Researchers in \cite{OPP} defined the OPP as the point at which the discriminability of spectral features between the two classes was maximized. Works \cite{preictalTransition, preictalConnectivity} have computed the preictal length using clustering methods from non-linear, univariate, multivariate, and connectivity EEG measures. Results showed that the OPP varied from 5 to 173 minutes before the onset, while average values ranged from 25 to 48 minutes. As for the post-training approaches, the performance of models trained with varying preictal lengths -- usually ranging from 5 to 120 minutes -- is evaluated against predefined metrics, such as accuracy, sensitivity, specificity, and F1-score, with the most successful model being selected \cite{Khan, timefrequency, LSTM,CNNGRUAttention, varyingPreictal1}. Most studies observed higher classification accuracy with shorter preictal lengths, ranging from 5 to 10 minutes. 

Although the aforementioned strategies have helped mitigate the uncertainty in defining the preictal period, they still exhibit considerable limitations. Pre-training approaches are constrained by the quality and relevance of the hand-crafted EEG markers, which may not accurately represent the underlying dynamics of preictal activity. They are also agnostic of the histopathology of specific epilepsy subtypes and the location of the epileptogenic zone. Post-training methods, though less sophisticated, demonstrate greater resilience to our incomplete understanding of preictal state dynamics and align more directly with the targeted diagnostic task. Nonetheless, conventional metrics used for model comparison fail to provide a comprehensive assessment of system performance. Though indicative of the model's ability to distinguish preictal from interictal segments, they capture neither the classifier's behavior during the gradual transition to the preictal state nor the distribution of false positives and negatives, which are critical for evaluating the model's practical implementation. Few studies \cite{longEEG, longEEG2, zhang_2016} have employed continuous, long-term EEG to visualize the system's behavior, yet they still rely on standard metrics without introducing novel measures. 

\subsection{Contributions}
\label{sub:introContributions}
Recognizing these gaps, this study introduces a novel methodology for evaluating epileptic seizure prediction models. By training a high-performing classifier, we aim to provide a comprehensive assessment that not only monitors the model’s behavior but also facilitates the selection of an optimal preictal state definition. For this, the contributions of this work are:
\begin{enumerate}
    \item We present a CNN-Transformer model to classify spatiotemporal EEG dynamics of preictal versus interictal EEGs that shows high sensitivity, early prediction time, and consistent performance across subjects. 
    \item We introduce an approach that allows nuance assessment of the deep learning model’s performance by fitting a sigmoidal curve to the output of the classifier subject to continuous, long EEG input. 
    \item We developed a novel Continuous Input-Output Performance Ratio (CIOPR) metric that provides a comprehensive assessment of the performance of a seizure prediction system in a realistic implementation setting by combining measures of prediction time, output stability, and transition time between interictal and preictal states. 
    \item We demonstrate the large impact of different preictal state definitions in the system’s performance as well as how the CIOPR metric can be utilized to determine the optimal preictal period for each patient (OPP). 
\end{enumerate}

\section{Materials and methods}
\label{sec:methods}
\subsection{EEG data and participant selection}
\label{sub:dataset}
% revisit this section 
The proposed seizure prediction model was trained and evaluated on the CHB-MIT dataset \cite{physionet}. It consists of scalp EEG recordings from 23 pediatric patients at the Children's Hospital Boston following an anti-seizure medication withdrawal period of seven days. Patients' ages ranged from 1.5 to 22 years, with individual epilepsy types not being mentioned. There were 198 recorded seizures in total, for which seizure start and end times were annotated by experts. Cases \textit{chb01} and \textit{chb21} came from the same subject, with the latter being recorded 1.5 years later. Subject information for case \textit{chb24} was not provided. Electrodes were placed according to the international 10-20 system with the number of channels for most of the patients ranging from 23 to 26. The sampling rate was set to 256 Hz with 16-bit resolution.

Data recorded up to 4 hours before and 1 hour after each annotated seizure were classified as interictal, to account for the uncertainty in the duration of preictal and postictal effects. Cases without the defined criteria for interictal data were deemed ineligible and excluded from the study, regardless of the number of recorded seizures. The maximum duration for the preictal period was set to 60 minutes. Similarly to the interictal class, data collected up to 1 hour after seizure termination were not considered preictal. Seizures yielding less than one minute of preictal data were excluded as ineligible. Additionally, cases with fewer than two eligible seizures were also excluded, regardless of the volume of available interictal data, since there should be at least one seizure for training and one for testing each subject-specific model. Based on the exclusion criteria mentioned above, cases \textit{chb08}, \textit{chb12}, \textit{chb13}, \textit{chb15}, and \textit{chb24} were excluded. Cases \textit{chb01} and \textit{chb21} were treated as separate subjects due to the long time difference between the recordings. In instances where electrode placement varied across recordings, the configuration yielding the greatest amount of data was selected. A summary of all patient data and eligible seizures, as well as their \textit{Patient ID}, \textit{Seizure ID}, and file name, are detailed in the Supplementary Material Tables 1 and 2, respectively. 

\subsection{Pre-processing}
\label{sub:pre-processing}
Pre-processing of the EEG signals was kept minimal and was performed on the files meeting the eligibility criteria in Section~\ref{sub:dataset}. Firstly, EEG was re-referenced to common average, which has been shown to improve the Signal-to-Noise Ratio (SNR) \cite{eeg_detection}. A 0.5 to 45 Hz Finite Impulse Response (FIR) non-causal bandpass filter with Hamming window and order $N=1690$ was designed in the time-domain to remove redundant frequency bands, using the MNE Python package. The high-pass cutoff at 0.5 Hz served to remove DC offset and low-frequency voltage drifts \cite{drift}, while the low-pass cutoff at 45 Hz removed power line interference and additional scalp EEG artifacts present at higher frequencies. No channels were removed, resulting in 23 channels for each eligible patient. Pre-processed data were then segmented into 5-second non-overlapping epochs. 

\subsection{Training methodology}
\label{sub:training}
Four different preictal period lengths -- 60, 45, 30, and 15 minutes -- were then extracted from each seizure to explore their effect on model performance and determine the OPP for each subject separately. Data extraction was conducted as follows: for the 60-minute duration, all available preictal data were extracted for each seizure, regardless of the actual duration available (e.g., some seizures had only 20 minutes of data available). For the 45-minute duration, only segments from the 45 minutes immediately preceding the seizure onset were included; if a seizure had 60 minutes of preictal data available, the first 15 minutes were excluded. If, for example, only 20 minutes were available, none were excluded. This procedure was similarly applied to the 30-minute and 15-minute durations. A histogram of available preictal data per seizure (in minutes) can be found in the Supplementary Material, Figure 1. 

The models were trained and evaluated in a subject-specific manner. For each definition of preictal length, the extracted preictal segments from all seizures were concatenated without shuffling. To mitigate class imbalance, the minority class (preictal) data were oversampled by a factor of 3 through the introduction of a 66\% overlap. Similarly, interictal segments were randomly selected from the pool of interictal data to match the number of the augmented preictal data instances. The training and testing sets were created using the Leave-One-Seizure-Out Cross-Validation (LOOCV) method. Specifically, preictal data of a different seizure each time were isolated, alongside an equal number of randomly selected interictal segments to form the testing set. The remaining samples were shuffled and divided into a 90\% training and 10\% validation split, forming the training set. This process was repeated four times to ensure consistency across different shuffling and random weight initializations. Ultimately, this resulted in $N \times 4 \times k \times 4$ training and testing sets, where $N$ represents the number of eligible subjects, $k$ denotes the number of eligible seizures per patient, and the four iterations correspond to the different preictal lengths (60, 45, 30, and 15 minutes), as well as the number of different runs.

\subsection{Deep learning model}

\label{sub:training}
\begin{figure*}[t]
\centering
\includegraphics[width=\textwidth]{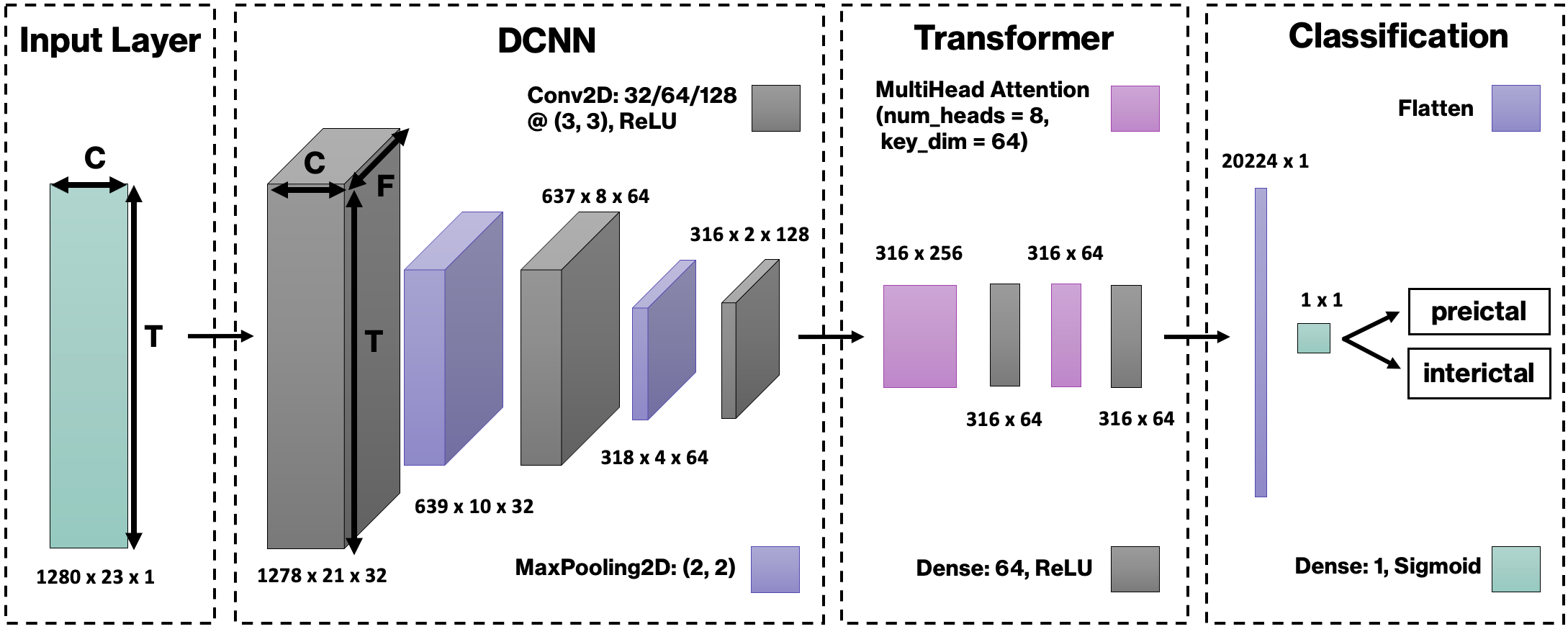}
\caption{\small Architecture of the CNN-Transformer deep learning model. Inputs were of shape $1280\times 23$, for 5-second 23-channel EEG segments. Dimensions $T$, $C$, and $F$ correspond to time points, channels, and feature maps respectively. The shape of the data following each layer is displayed as $T\times C \times F$. For the transformer stage, $C$ and $F$ are aggregated into a single dimension and expressed as $T\times (C, F)$. Specific information on each layer can be found in the Supplementary Material, Table 3.}
\label{fig:deeplearning}
\end{figure*}

A CNN-Transformer deep learning model was developed to classify preictal and interictal EEG segments, with the architecture illustrated in Figure~\ref{fig:deeplearning}. The raw pre-processed EEG epochs were first input to a three-layer Deep Convolutional Neural Network (DCNN) to extract relevant spatiotemporal features. Each convolutional layer, consisting of 32, 64, and 128 filters, respectively, applied a $3\times3$ filter and was followed by a Max-Pooling operation that reduced dimensionality by a factor of two. The architecture also incorporated a Dropout layer with a probability of $p=0.1$ before each Max-Pooling step to mitigate overfitting, complemented by Batch Normalization (BN) and a Rectified Linear Unit (ReLU) activation function.

After the final convolutional layer, the output tensor was reshaped into a two-dimensional matrix, where rows corresponded to subsequent time points and columns to spatiotemporal features, including positional information resulting from integrating the channel and feature dimensions during the reshaping process. To capture long-term temporal dependencies among these features, a transformer architecture was utilized, with two multi-head attention layers \cite{attention}. The reshaped two-dimensional output was used for the creation of the query, key, and value matrices. The linear layers shrunk the dimension of the feature vectors to 64, while the number of attention heads was set to 8. Each attention layer was followed by a fully-connected (dense) layer of 64 neurons, ReLU activation and $p=0.3$ dropout, to highlight the most relevant features, while maintaining the dimensionality of the feature vectors. 

The resulting attention-weighted feature map was then flattened and processed through a single-neuron output layer with sigmoid activation for the classification stage. The output value was rounded to the nearest integer (0: interictal, 1: preictal) without further post-processing. The model was trained using the binary cross-entropy loss function \cite{lossfunctions}, with weight updates performed via the Adam optimization algorithm. The learning rate was set to $l=0.001$, and the AMSGrad \cite{optimizers} extension was enabled. The training utilized a mini-batch size of 64 and was limited to 100 epochs. To prevent overfitting, the early-stopping callback terminated training if the validation loss did not improve for 20 consecutive epochs. The model-checkpoint callback preserved the weights of the model iteration, achieving the lowest validation loss. The complete summary of the model can be found in the Supplementary Material. 

\subsection{Continuous Input-Output Performance Ratio (CIOPR)}
\label{sub:methodsCIOPR}
We introduce a novel metric that facilitates direct comparison of classifier behaviors across prediction tasks. Traditional accuracy metrics, heavily reliant on class definitions, fail to capture the timing of predictions or model performance during state transitions. Addressing these limitations, our innovative metric, CIOPR enables a comprehensive evaluation of models using uninterrupted, unlabeled long-term EEG data as input, to establish objective comparisons independent of class definitions. We utilized it to compare the effect of different preictal state definitions and assumed that the ideal classifier would offer early prediction, minimal errors, high stability, and brief transition periods.

When subjected to continuous EEG data spanning several hours before a seizure, an accurate classifier is anticipated to initially produce negative (interictal) predictions, transition through a mix of negative and positive predictions, and ultimately converge to the preictal state. Consequently, to quantitatively model the timing of these predictions, we propose to fit a sigmoidal curve to the classifier’s continuous output, which has undergone smoothing with an 8-minute averaging window. In particular, a 4-Parameter Logistic curve (4PL) is used, given by Equation~\ref{eq:sigmoid} \cite{sigmoid}, where parameters $a$, $b$, $c$, and $d$ represent the vertical and horizontal stretch, the point of inflection, and the vertical offset, respectively, as
\begin{equation}
\label{eq:sigmoid}
f(x) = \frac{a}{1 + e^{-b(x-c)}} + d.
\end{equation}

\begin{figure}[!htb]
\centering
\includegraphics[width=0.5\textwidth]{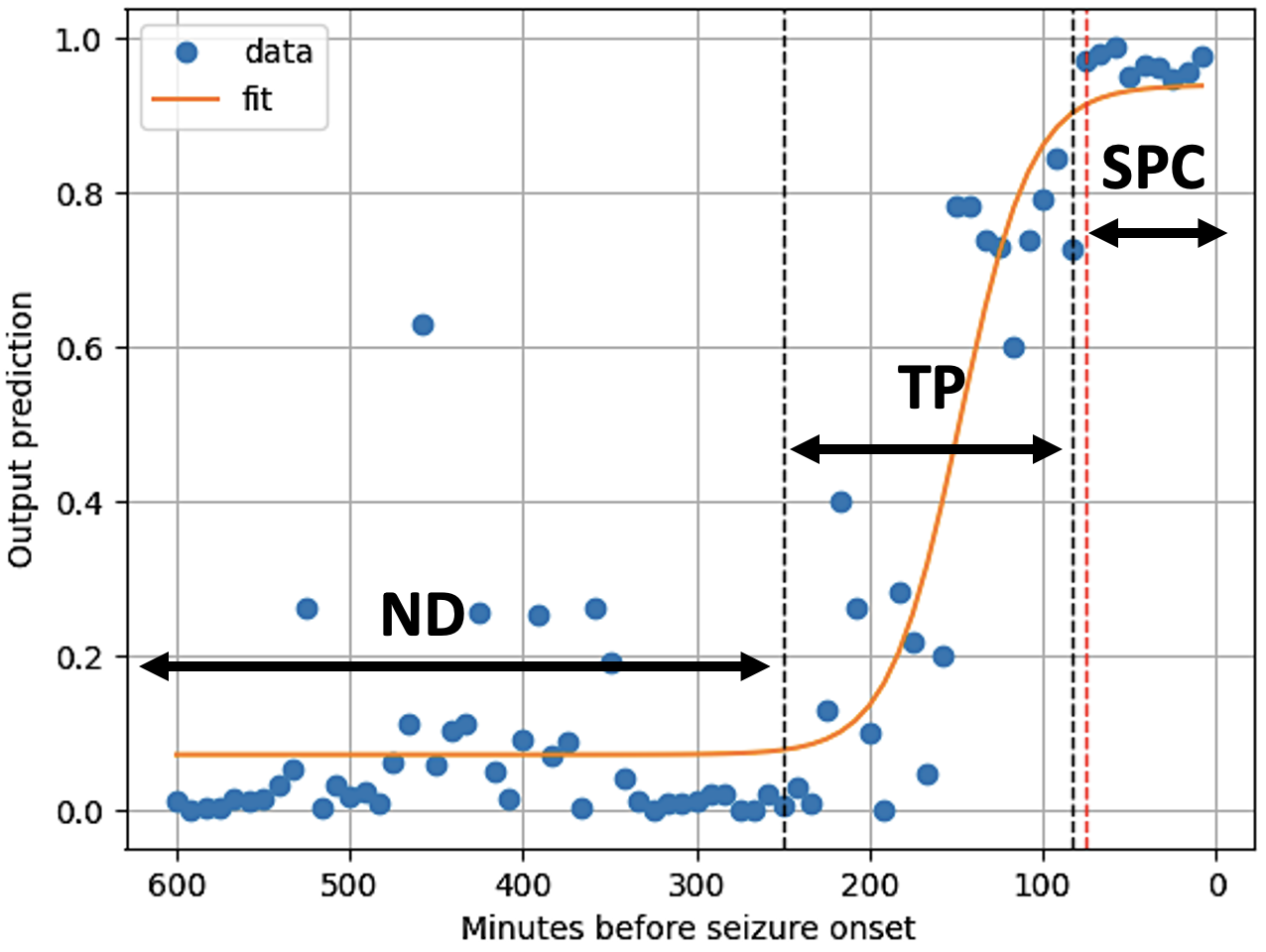}
\caption{\small Continuous output predictions based on 600 minutes of EEG data preceding seizure onset, along with the fitted sigmoidal curve. The black dotted vertical lines, derived from the sigmoid fit, denote the start and end points of the transition period between interictal and preictal states. The red dotted line, directly calculated from the output, designates the start of the seizure prediction convergence, where the classifier's output converges. Notations: $ND$ = Negative Duration (interictal predictions), $TP$ = Transition Period, and $SPC$ = Seizure Prediction Convergence.
}
\label{fig:ciopr}
\end{figure}

Utilizing the fitted sigmoidal curve, we derive key performance measures. Specifically, the transition period ($TP$) between interictal and preictal states is quantified as the time interval between the 5th and 95th percentile thresholds of the sigmoid curve's amplitude. The negative duration ($ND$), the supposed interictal period, is calculated as the duration between the first output prediction and the beginning of the transition period. The remaining measures are directly computed from the output data, following average smoothing (1 output prediction every 8 minutes). The seizure prediction convergence ($SPC$) is the point in time where output predictions reached 99\% of the maximum value, in minutes before seizure onset. It represents the period of highest incoming-seizure probability by the model and could reflect the prediction horizon in a realistic implementation setting. Both output and maximum values used for the computation of $SPC$ refer to the mean of 3 consecutive predictions ($3\times8~minutes=24~minutes$).
Figure \ref{fig:ciopr} depicts the fitted sigmoid, the output predictions, and the aforementioned measures. 

The average errors during the $SPC$ ($SPC_{err}$) and interictal period ($ND_{err}$) are calculated using Equations \ref{eq:shperr} and \ref{eq:nderr}, respectively. $N_{SPC}$ and $N_{ND}$ refer to the number of output predictions in each of the two regions. Then, Equation \ref{eq:ciopc} is developed to combine all the measures so that the classifier with the longest $SPC$, minimum $SPC_{err}$ and $ND_{err}$, and shortest $TP$ achieved the highest score. $SPC_{eff}$ and $ND_{eff}$ represent the effective values of $SPC$ and $ND$ taking into account the average errors, and defined as $SPC_{eff}=SPC(1-SPC_{err})$ and $ND_{eff}=ND(1-ND_{err})$.

\begin{equation}
SPC_{err} = \frac{1}{N_{SPC}} \sum_{i=1}^{N_{SPC}} |1 - y_i|
\label{eq:shperr}
\end{equation}

\begin{equation}
ND_{err} = \frac{1}{N_{ND}} \sum_{i=1}^{N_{ND}} |y_i|
\label{eq:nderr}
\end{equation}

\begin{equation}
CIOPC = SPC_{eff} + \eta (ND_{eff} + Infl_{comp}) \\
\label{eq:ciopc}
\end{equation}

Overall, Equation \ref{eq:ciopc} computes the Continuous Input-Output Performance Coefficient (CIOPC) of a model by taking into account two terms. The $SPC_{eff}$ term rewards the model for early prediction ($SPC$), as well as high accuracy ($SPC_{err}$). As for the second term, $ND_{eff}$ rewards the model for short $TP$ (longer $ND$ when $TP$ is short) and high accuracy ($ND_{err}$). A reduction in $ND$ due to earlier prediction should not get penalized and is compensated by the $Infl_{comp}$ term ($ND_{eff}$ lost due to an earlier point of inflection). The scaling factor $\eta$ ensures that the second term will not have a greater weighting than $SPC_{eff}$ due to hours-long interictal EEG; hence, maintaining the focus on early and accurate seizure prediction.

The CIOPC value for each preictal state definition is normalized to the CIOPR metric using Equation \ref{eq:ciopr}, where $CIOPC_{max}$ denotes the maximal CIOPC obtained across the examined preictal durations. This normalization facilitates a relative performance evaluation, where the highest-scoring model is assigned a CIOPR score of 1, thus enabling a comparative analysis of the impact of varying preictal period lengths. Detailed expressions for $Infl_{comp}$ and $\eta$, as well as elucidation of the underlying intuition for CIOPR, are detailed in the Supplementary Material, Section 3. 

\begin{equation}
\begin{split}
CIOPR_k = \frac{CIOPC_k}{CIOPC_{max}}, k \in \{60, 45, 30, 15\}
\label{eq:ciopr}
\end{split}
\end{equation}

Lastly, the Pearson correlation coefficient, $\rho$ is calculated between the output predictions and the fitted curve to assess the method's reliability. It also serves as an indicator of output stability, since greater fluctuations are expected to decrease the correlation coefficient. 

\subsection{Additional metrics}
\label{sub:additionalmetrics}
Beyond the CIOPR, conventional performance metrics including segment-wise sensitivity (SEN), specificity (SPE), accuracy (ACC), and F1-score (F1) were employed for performance evaluation. These metrics were computed on the test set, the formulation of which is detailed in Section \ref{sub:training}. Given the equal representation of classes in the test set, the resulting accuracy is equivalent to the balanced accuracy.

\subsection{Testing setup}
\label{sub:testing}
For each eligible seizure of a qualifying participant (criteria detailed in Section \ref{sub:dataset}), the prediction performance was tested using the conventional metrics (SEN, SPE, ACC, and F1), for all preictal durations. Seizures with over 2.5 hours of uninterrupted continuous EEG data were further subjected to CIOPR evaluation for each preictal interval. A maximum of 10 hours of continuous EEG was utilized for CIOPR to reduce computation time and maintain consistency across patients. 

Metrics were then aggregated across seizures to compute the average performance for each preictal duration. For conventional metrics, averages were calculated on a segment-wise basis rather than seizure-wise, ensuring that testing seizures with more data samples were accorded greater weight. These averages were then used to determine the optimal preictal period (OPP). Table 1 of the Supplementary Material provides a summary of the recorded seizures per subject, including both the count of eligible seizures and those applicable for CIOPR testing.

\subsection{OPP selection}
A general OPP -- 60, 45, 30, or 15 minutes -- was then selected for each participant. When at least one seizure was eligible for CIOPR testing, the OPP was assigned to the preictal definition yielding the highest CIOPR values, averaged across those seizures. Conversely, if no CIOPR results were available, the OPP was assigned to the preictal definition attaining the highest F1-score, averaged across all testing seizures. F1-score is preferred over balanced accuracy due to its capability to emphasize minimizing false negatives, which are considered more detrimental in epileptic seizure prediction \cite{DLEpilepsy}. Lastly, the model trained using the OPP was selected from each patient to evaluate both the subject-specific and aggregate performance of our proposed methodology. To further showcase performance, the False Alarm Rate (FAR) ($h^{-1}$) and the Area Under the Receiver Operating Characteristic (ROC) Curve (AUC) were reported only for the selected models. FAR was computed based on the EPOCH \cite{EPOCH} method on the 5-second segment level.   

\subsection{Statistical analysis}
\label{sub:statistical_analysis}
To explore statistical differences in model behavior attributed to different preictal definitions, a related-samples Friedman's two-way ANOVA with Bonferroni correction for multiple comparisons was conducted for all the metrics used in the testing setup ($SEN$, $SPE$, $ACC$, F1-score, $SPC$, $SPC_{err}$, $ND_{err}$, inflection point, $TP$, Pearson correlation, $CIOPR$). The null hypothesis tested was that the distributions of scores for each metric across the preictal lengths of 60, 45, 30, and 15 minutes do not differ significantly (significance level $\alpha=0.05$ after correction). 

\section{Results}
\label{sec:results}
%Structure of results section: 
%Show that for several patients accuracies were similar but performance was very different for each preictal definition.
%Include table that shows all desired metrics as well as box-plots illustrating differences for each preictal duration.
%The below figure is for testing purposes. Example plot supporting the table:

%Other patients exhibit slightly better behaviour with slightly better accuracy. I would recommend grouping the results like this to allow for more readable tables. In the end insert a table with all patients but only on the selected preictal definition per patient. 

\subsection{Classifier Training}

The models were trained using an A100 Tensor Core GPU in Google Colab. The average training duration was 8.1 minutes per model, converging at epoch 54 due to the early-stopping callback. The training duration was significantly longer for 60- and 45-minute preictal class definitions due to increased data volume. The convergence minima for both validation and training curves varied more across patients than across intra-patient seizures or different preictal state lengths.

\subsection{Classification results}
\label{sub:f1}

Table \ref{tab:test1_table} details the segment-wise $SEN$, $SPE$, $ACC$, and F1-score of the subject-specific classifiers. The displayed values represent the weighted average across all runs and testing seizures for each patient. A preictal class duration of 60 minutes generally yielded the highest average F1-scores, although 45- and 30-minute definitions performed better in certain individuals. The percentage change in F1-scores across different preictal class definitions was relatively modest, with a maximum variation of $3.2\%$ observed in case \textit{chb05}. Sensitivity scores remained consistent across various preictal durations. Fluctuations in the F1-score were primarily attributable to reductions in specificity associated with shorter preictal states. 

\begin{table*}[htbp]
\centering
\caption{Subject-Specific Performance based on Conventional Metrics}
\label{tab:test1_table}
\begin{adjustbox}{max width=\textwidth}
\begin{tabular}{@{}lcccccccccccccccc@{}}
\toprule
Case & \multicolumn{4}{c}{Sensitivity (\%)} & \multicolumn{4}{c}{Specificity (\%)} & \multicolumn{4}{c}{Accuracy (\%)} & \multicolumn{4}{c}{F1-Score (\%)} \\ \cmidrule(lr){2-5} \cmidrule(lr){6-9} \cmidrule(lr){10-13} \cmidrule(l){14-17} 
 & 60 mins & 45 mins & 30 mins & 15 mins & 60 mins & 45 mins & 30 mins & 15 mins & 60 mins & 45 mins & 30 mins & 15 mins & 60 mins & 45 mins & 30 mins & 15 mins \\ \midrule
chb01 & 100.00 & 100.00 & 99.96 & 100.00 & 99.85 & 99.85 & 99.82 & 99.47 & 99.93 & 99.93 & 99.89 & 99.74 & 99.93 & 99.93 & 99.89 & 99.73 \\
chb02 & 99.76 & 99.75 & 100.00 & 99.70 & 96.32 & 96.68 & 95.08 & 94.92 & 98.04 & 98.22 & 97.54 & 97.31 & 98.08 & 98.24 & 97.59 & 97.36 \\
chb03 & 99.80 & 99.80 & 99.87 & 99.45 & 96.91 & 96.91 & 97.56 & 97.35 & 98.36 & 98.36 & 98.72 & 98.40 & 98.38 & 98.38 & 98.73 & 98.41 \\
chb04 & 99.96 & 99.84 & 99.87 & 99.61 & 99.43 & 99.19 & 99.10 & 99.07 & 99.70 & 99.52 & 99.49 & 99.34 & 99.70 & 99.51 & 99.49 & 99.34 \\
chb05 & 99.62 & 99.62 & 99.55 & 99.04 & 96.66 & 95.59 & 94.17 & 92.60 & 98.14 & 97.61 & 96.86 & 95.82 & 98.16 & 97.66 & 96.95 & 95.95 \\
chb06 & 94.76 & 94.85 & 97.10 & 97.43 & 73.54 & 75.59 & 75.14 & 73.39 & 84.15 & 85.22 & 86.12 & 85.41 & 85.71 & 85.22 & 86.12 & 85.41 \\
chb07 & 99.54 & 99.84 & 99.64 & 99.45 & 97.74 & 97.48 & 97.69 & 96.09 & 98.64 & 98.66 & 98.67 & 97.77 & 98.66 & 98.68 & 98.71 & 97.73 \\
chb09 & 99.72 & 99.29 & 99.62 & 99.87 & 98.85 & 97.75 & 96.74 & 94.09 & 99.29 & 98.52 & 98.18 & 96.98 & 99.29 & 98.55 & 98.14 & 97.21 \\
chb10 & 98.40 & 98.85 & 98.57 & 98.46 & 92.84 & 91.76 & 90.87 & 86.60 & 95.62 & 95.31 & 94.72 & 92.53 & 95.76 & 95.46 & 94.91 & 92.81 \\
chb11 & 99.86 & 99.86 & 100.00 & 100.00 & 99.81 & 99.81 & 99.32 & 95.71 & 99.84 & 99.84 & 99.66 & 97.86 & 99.83 & 99.83 & 99.65 & 97.90 \\
chb14 & 97.85 & 95.48 & 98.54 & 97.36 & 86.81 & 84.65 & 86.00 & 85.22 & 92.33 & 90.07 & 92.27 & 91.29 & 92.73 & 90.58 & 92.67 & 91.84 \\
chb16 & 99.64 & 99.78 & 99.41 & 99.06 & 97.00 & 94.16 & 94.31 & 92.20 & 98.32 & 96.97 & 96.86 & 95.63 & 98.33 & 96.94 & 96.94 & 95.75 \\
chb17 & 99.45 & 99.85 & 99.37 & 99.76 & 97.57 & 96.69 & 95.69 & 95.08 & 98.51 & 98.27 & 97.53 & 97.42 & 98.51 & 98.30 & 97.57 & 97.47 \\
chb18 & 99.59 & 99.24 & 99.73 & 99.16 & 96.40 & 95.55 & 94.75 & 95.61 & 98.00 & 97.40 & 97.24 & 97.39 & 98.02 & 97.43 & 97.32 & 97.43 \\
chb19 & 100.00 & 99.88 & 100.00 & 100.00 & 99.11 & 99.50 & 99.64 & 99.46 & 99.56 & 99.69 & 99.82 & 99.73 & 99.30 & 99.69 & 99.82 & 99.73 \\
chb20 & 99.84 & 99.85 & 99.96 & 100.00 & 97.66 & 98.05 & 98.05 & 98.84 & 98.75 & 98.95 & 99.01 & 99.42 & 98.64 & 98.85 & 99.01 & 99.42 \\
chb21 & 100.00 & 99.74 & 99.86 & 99.64 & 97.01 & 97.03 & 96.00 & 97.10 & 98.50 & 98.39 & 97.93 & 98.37 & 98.50 & 98.41 & 97.98 & 98.37 \\
chb22 & 99.24 & 99.19 & 99.71 & 99.16 & 87.45 & 86.38 & 85.70 & 82.93 & 93.35 & 92.79 & 92.71 & 91.05 & 93.77 & 93.23 & 93.11 & 91.74 \\
chb23 & 100.00 & 99.78 & 99.95 & 99.69 & 99.36 & 98.74 & 98.41 & 98.09 & 99.68 & 99.26 & 99.18 & 98.89 & 99.68 & 99.26 & 99.18 & 99.90 \\
\midrule
Mean & 99.28 & 99.14 & 99.49 & 99.27 & 95.03 & 94.53 & 94.12 & 93.02 & 97.15 & 96.83 & 96.80 & 96.14 & 97.28 & 96.90 & 96.88 & 96.32 \\
$\pm$ Std & 1.24 & 1.45 & 0.73 & 0.79 & 6.43 & 6.25 & 6.21 & 6.81 & 3.80 & 3.79 & 3.43 & 3.75 & 3.44 & 3.71 & 3.37 & 3.69 \\
\bottomrule
\end{tabular}
\end{adjustbox}
\end{table*}

\subsection{Continuous Input-Output Performance Ratio (CIOPR) Testing Results}
\label{sub:ciopr}

The proposed CIOPR in Section \ref{sub:methodsCIOPR} was performed on the thirteen patients that met the criteria described in Section \ref{sub:testing}. The first step of the analysis involved fitting the sigmoidal curve to the smoothed output of the classifiers. The Pearson correlation coefficient between the model output and the fitted curve across all the tested patients was on average $\rho>0.9$ for the 60, 45, and 30-minute preictal definitions, and $\rho=0.876$ for the 15-minute one. The high correlation values show the effectiveness of the chosen curve in modeling the classifiers' output profile and indicate that the models generally exhibited the intended behaviour. Similar to the F1-score results, the correlation coefficients differed across patients, seizures and preictal state lengths, with 60 minutes usually leading to higher values.

\begin{figure}[h]
        \centering
        \begin{subfigure}[b]{0.23\textwidth}
            \centering
            \includegraphics[width=\textwidth]{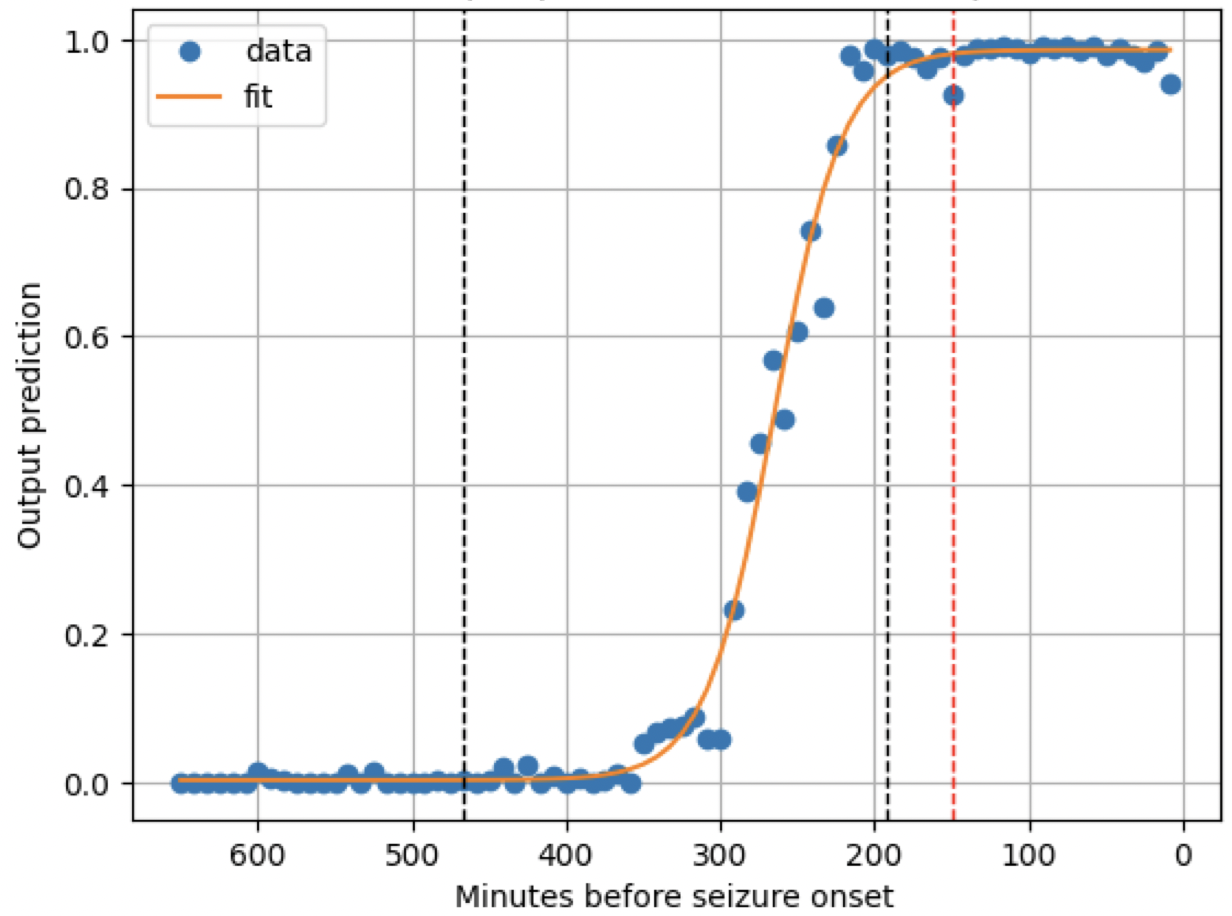}
            \caption[Network2]%
            {{\small Case chb21, 1st seizure, 45 mins, $\rho$ = 0.9959}}    
            \label{fig:corr_a}
        \end{subfigure}
        \hfill
        \begin{subfigure}[b]{0.23\textwidth}  
            \centering 
            \includegraphics[width=\textwidth]{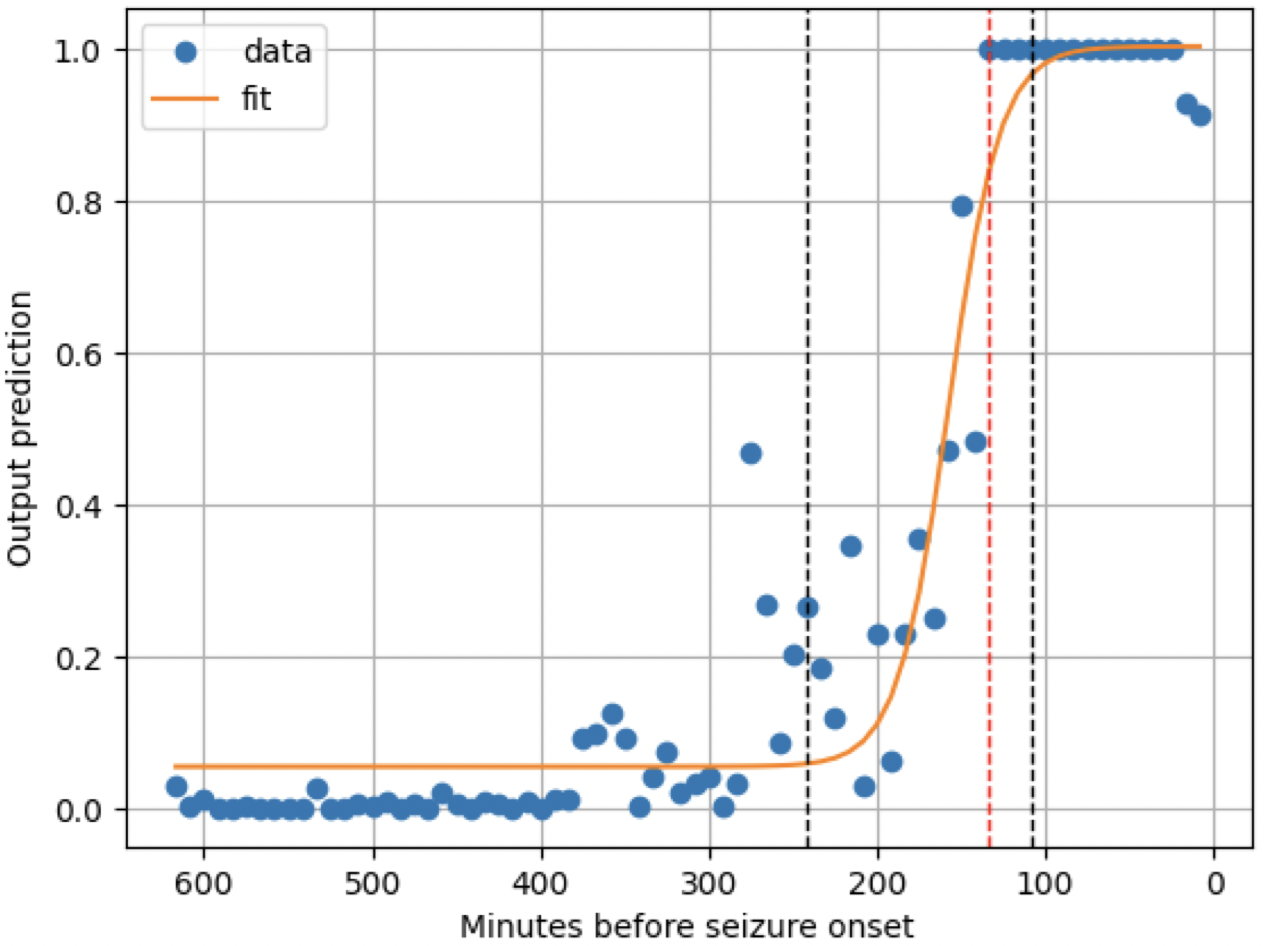}
            \caption[]%
            {{\small Case chb04, 1st seizure, 45 mins, $\rho$ = 0.9531}}    
            \label{fig:corr_b}
        \end{subfigure}
        \vskip\baselineskip
        \begin{subfigure}[b]{0.23\textwidth}   
            \centering 
            \includegraphics[width=\textwidth]{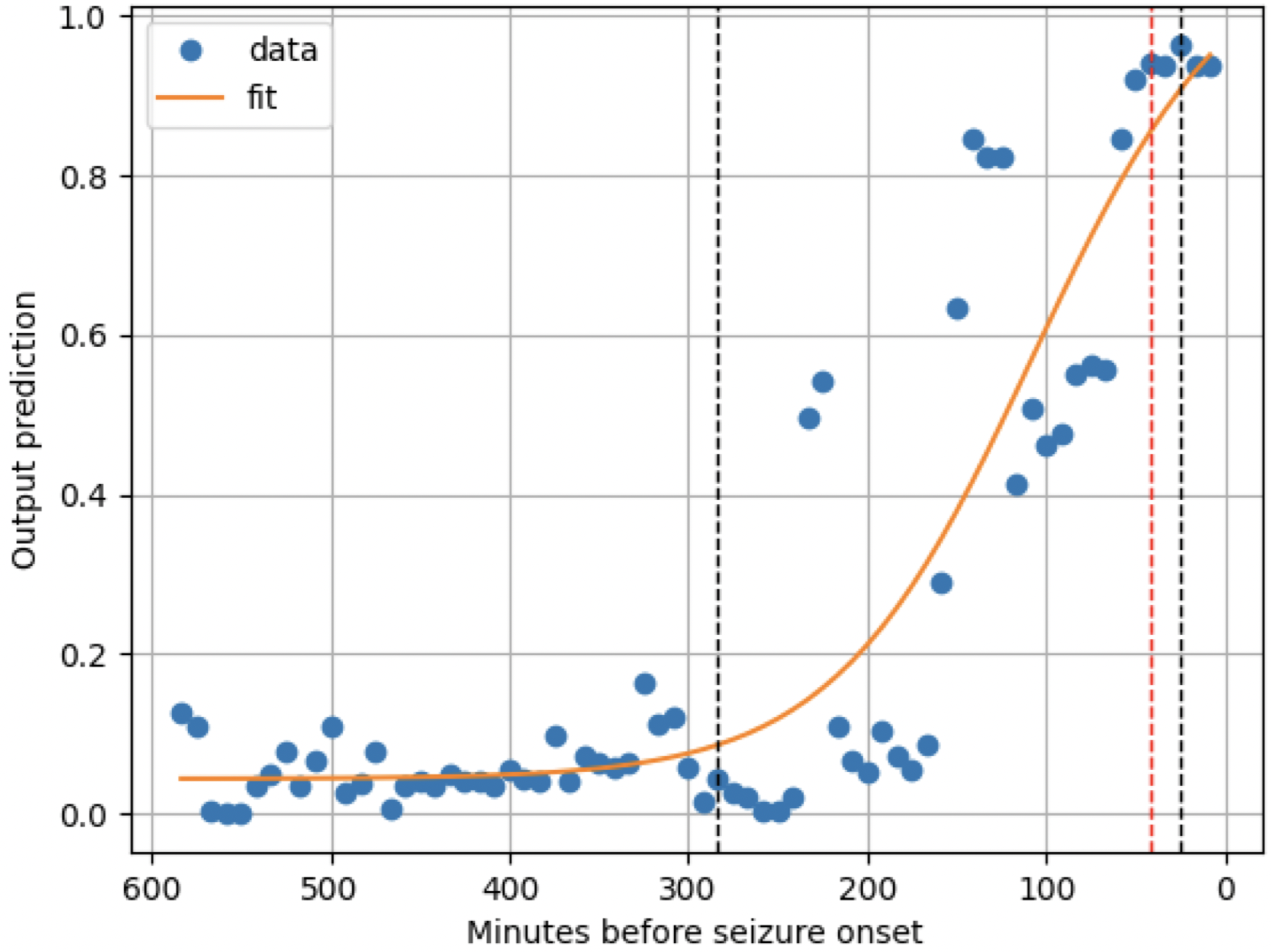}
            \caption[]%
            {{\small Case chb16, 1st seizure, 45 mins, $\rho$ = 0.8992}}    
            \label{fig:corr_c}
        \end{subfigure}
        \hfill
        \begin{subfigure}[b]{0.23\textwidth}   
            \centering 
            \includegraphics[width=\textwidth]{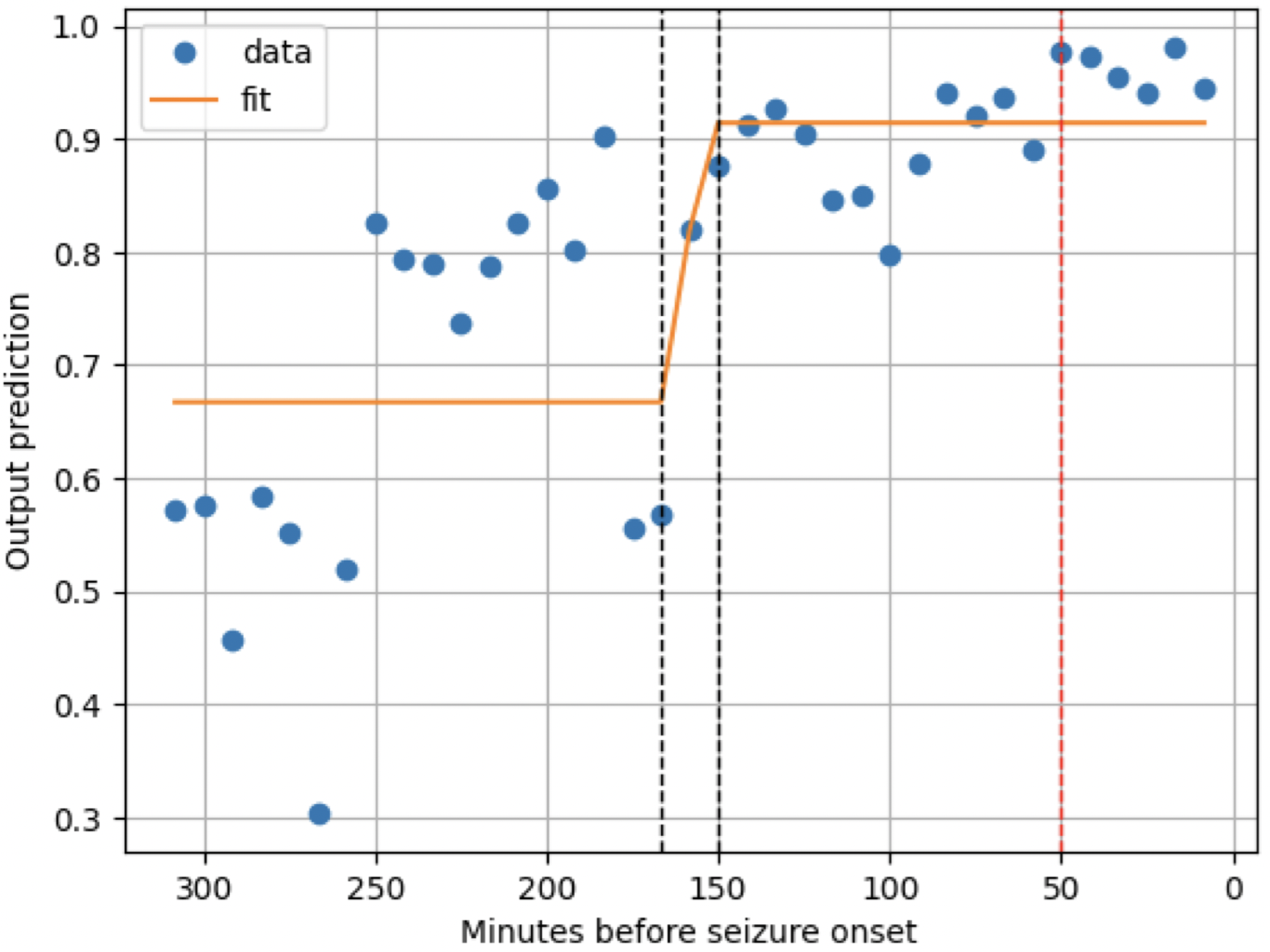}
            \caption[]%
            {{\small Case chb05, 1st seizure, 45 mins, $\rho$ = 0.7254}}    
            \label{fig:corr_d}
        \end{subfigure}
        \caption[ Correlations ]
        {\small Four figures showing different levels of fitting between the classifier's output and the fitted sigmoid curve, sorted with decreasing Pearson correlation coefficient. The horizontal axis represents minutes before seizure onset (seizure onset at utmost right), and the vertical axis is the classification output. Captions show the case number, seizure ID, preictal class definition, and Pearson correlation coefficient.} 
        \label{fig:correlations}
    \end{figure}

Figure \ref{fig:correlations} demonstrates four fitting examples to allow visualization of different correlation coefficients and corresponding model behaviour. Sub-figures \ref{fig:corr_a} and \ref{fig:corr_b} display high correlation coefficients ($>0.95$). Sub-figure \ref{fig:corr_c} depicts a fitting with a correlation coefficient of approximately $0.9$, enabling reliable identification of both the prediction horizon and the transition period. Conversely, sub-figure \ref{fig:corr_d} shows one of the fittings with the lowest correlation ($<0.75$). While the transition period and interictal state were less discernible, a noticeable shift in the density of positive predictions still enabled the identification of the preictal state start and subsequent performance quantification. Seizure 2 of case \textit{chb05} was the only instance in which the curve could not be fitted and was therefore excluded from the analysis. The correlation coefficients were generally reduced with decreasing preictal length, primarily due to the increased volume of ``unseen'' data from the classifiers. 
A detailed list of average correlation coefficients for all cases can be found in the Supplementary Material, Table 4.

The fitted curves allowed the computation of the CIOPR values for each testing seizure per patient, which were then averaged and used to determine the optimal preictal class definition. Table \ref{tab:allciopr} presents the CIOPR and corresponding F1-score values for each eligible seizure and patient. Notably, the F1-scores in Table \ref{tab:allciopr} are averaged only from seizures used for CIOPR testing, unlike those in Table \ref{tab:test1_table}, which include all testing seizures per case. Additionally, the table displays average SPC values for each patient, illustrating the prediction time achieved with each preictal class definition. Detailed values of all metrics used to compute the CIOPR scores are available in the Supplementary Material, Table 4.

\begin{table*}[htbp]
\centering
\caption{SPC, CIOPR and F1-score results for all testing seizures applicable for CIOPR analysis}
\label{tab:allciopr}
\begin{adjustbox}{max width=\textwidth}
\begin{tabular}{@{}lcccccccccccccccc@{}}
\toprule
Case & Seiz. ID & \multicolumn{4}{c}{SPC (mins)} & \multicolumn{4}{c}{CIOPR} & \multicolumn{4}{c}{F1-score}\\ \cmidrule(lr){3-6} \cmidrule(lr){7-10} \cmidrule(l){11-14}
& &  60 mins & 45 mins & 30 mins & 15 mins & 60 mins & 45 mins & 30 mins & 15 mins & 60 mins & 45 mins & 30 mins & 15 mins \\
\midrule
chb01 & 1 & 72.2 & 72.9 & 72.9 & 75.0 & 0.9582 & 0.9739 & 0.9806 & 1.0000 & 0.9985 & 0.9982 & 0.9981 & 0.9940 \\
chb01 & 2 & 66.7 & 62.5 & 64.6 & 50.0 & 1.0000 & 0.9803 & 0.9836 & 0.8954 & 0.9992 & 0.9989 & 0.9977 & 0.9962 \\
chb01 & 5 & 54.2 & 50.0 & 33.3 & 33.3 & 0.9763 & 1.0000 & 0.8348 & 0.8558 & 0.9982 & 0.9982 & 0.9958 & 0.9951 \\
chb01 & Avg & 64.3 & 61.8 & 56.9 & 52.8 & 0.9782 & \textbf{0.9847} & 0.9330 & 0.9171 & \textbf{0.9986} & 0.9984 & 0.9972 & 0.9951 \\
\midrule
chb02 & 1 & 66.7 & 52.1 & 41.7 & 33.3 & 1.0000 & 0.8924 & 0.8211 & 0.6644 & 0.9772 & 0.9753 & 0.9673 & 0.9497 \\
chb02 & Avg & 66.7 & 52.1 & 41.7 & 33.3 & \textbf{1.0000} & 0.8924 & 0.8211 & 0.6644 & \textbf{0.9772} & 0.9753 & 0.9673 & 0.9497 \\
\midrule
chb04 & 1 & 133.3 & 133.3 & 133.3 & 133.3 & 0.9810 & 1.0000 & 0.9873 & 0.9676 & 0.9969 & 0.9940 & 0.9960 & 0.9931 \\
chb04 & 2 & 208.3 & 217.6 & 208.3 & 166.7 & 0.9578 & 1.0000 & 0.9558 & 0.7671 & 0.9960 & 0.9957 & 0.9945 & 0.9947 \\
chb04 & Avg & 170.8 & 175.5 & 170.8 & 150.0 & 0.9694 & \textbf{1.0000} & 0.9716 & 0.8674 & \textbf{0.9965} & 0.9949 & 0.9952 & 0.9939 \\
\midrule
chb05 & 1 & 62.5 & 58.3 & 41.7 & 33.3 & 1.0000 & 0.9038 & 0.7412 & 0.7059 & 0.9832 & 0.9795 & 0.9738 & 0.9595 \\
chb05 & 4 & 47.2 & 41.7 & 36.1 & 25.0 & 1.0000 & 0.9150 & 0.7881 & 0.5814 & 0.9838 & 0.9776 & 0.9670 & 0.9592 \\
chb05 & Avg & 54.9 & 50.0 & 38.9 & 29.2 & \textbf{1.0000} & 0.9094 & 0.7647 & 0.6437 & \textbf{0.9835} & 0.9786 & 0.9670 & 0.9593 \\
\midrule
chb06 & 4 & 36.1 & 36.1 & 25.0 & 16.7 & 0.9912 & 1.0000 & 0.8903 & 0.7516 & 0.8536 & 0.8465 & 0.8737 & 0.8535 \\
chb06 & 6 & 52.8 & 47.2 & 33.3 & 25.0 & 1.0000 & 0.9492 & 0.8054 & 0.7139 & 0.8509 & 0.8632 & 0.8752 & 0.8650 \\
chb06 & 7 & 58.3 & 33.3 & 33.3 & 16.7 & 1.0000 & 0.6542 & 0.6705 & 0.4879 & 0.8491 & 0.8587 & 0.8421 & 0.8365 \\
chb06 & 8 & 66.7 & 41.7 & 30.6 & 25.0 & 1.0000 & 0.6257 & 0.4944 & 0.4700 & 0.8645 & 0.8523 & 0.8723 & 0.8700 \\
chb06 & 9 & 53.5 & 39.6 & 30.6 & 20.8 & 1.0000 & 0.8539 & 0.7727 & 0.6791 & 0.8881 & 0.8742 & 0.8812 & 0.8711 \\
chb06 & 10 & 50.0 & 41.7 & 33.3 & 16.7 & 0.9842 & 1.0000 & 0.9218 & 0.6936 & 0.8514 & 0.8725 & 0.8859 & 0.8663 \\
chb06 & Avg & 52.9 & 39.9 & 31.0 & 20.1 & \textbf{0.9959} & 0.8472 & 0.7592 & 0.6327 & 0.8596 & 0.8612 & \textbf{0.8717} & 0.8604 \\
\midrule
chb07 & 1 & 57.5 & 50.0 & 43.3 & 36.7 & 1.0000 & 0.9508 & 0.8871 & 0.8158 & 0.9861 & 0.9883 & 0.9881 & 0.9867 \\
chb07 & 2 & 66.7 & 50.0 & 41.7 & 33.3 & 1.0000 & 0.8598 & 0.7559 & 0.7193 & 0.9899 & 0.9882 & 0.9872 & 0.9676 \\
chb07 & 3 & 60.4 & 58.3 & 33.3 & 25.0 & 1.0000 & 0.8893 & 0.7332 & 0.7307 & 0.9838 & 0.9839 & 0.9860 & 0.9777 \\
chb07 & Avg & 61.5 & 52.8 & 39.4 & 31.7 & \textbf{1.0000} & 0.9000 & 0.7921 & 0.7553 & 0.9866 & 0.9868 & \textbf{0.9871} & 0.9773 \\
\midrule
chb09 & 1 & 125.0 & 97.2 & 66.7 & 41.7 & 1.0000 & 0.8850 & 0.7448 & 0.5701 & 0.9912 & 0.9871 & 0.9850 & 0.9671 \\
chb09 & 2 & 91.7 & 83.3 & 83.3 & 83.3 & 1.0000 & 0.9715 & 0.9022 & 0.8814 & 0.9918 & 0.9915 & 0.9726 & 0.9673 \\
chb09 & 4 & 63.9 & 63.9 & 41.7 & 25.0 & 0.9942 & 1.0000 & 0.4405 & 0.6925 & 0.9914 & 0.9817 & 0.9817 & 0.9736 \\
chb09 & Avg & 93.5 & 81.5 & 63.9 & 50.0 & \textbf{0.9981} & 0.9522 & 0.6958 & 0.7147 & \textbf{0.9915} & 0.9868 & 0.9798 & 0.9693 \\
\midrule
chb10 & 2 & 70.8 & 62.5 & 43.8 & 25.0 & 1.0000 & 0.9628 & 0.8035 & 0.6315 & 0.9623 & 0.9559 & 0.9572 & 0.9145 \\
chb10 & Avg & 70.8 & 62.5 & 43.8 & 25.0 & \textbf{1.0000} & 0.9628 & 0.8035 & 0.6315 & \textbf{0.9623} & 0.9559 & 0.9572 & 0.9145 \\
\midrule
chb14 & 1 & 50.0 & 41.7 & 33.3 & 16.7 & 1.0000 & 0.7731 & 0.6955 & 0.4538 & 0.9213 & 0.9119 & 0.9385 & 0.9286 \\
chb14 & 5 & 54.2 & 50.0 & 33.3 & 25.0 & 1.0000 & 0.9240 & 0.7424 & 0.6345 & 0.9289 & 0.9011 & 0.9169 & 0.9308 \\
chb14 & Avg & 52.1 & 45.8 & 33.3 & 20.8 & \textbf{1.0000} & 0.8486 & 0.7190 & 0.5442 & 0.9251 & 0.9065 & 0.9277 & \textbf{0.9297} \\
\midrule
chb16 & 1 & 62.5 & 43.8 & 33.3 & 41.7 & 1.0000 & 0.8215 & 0.7237 & 0.7261 & 0.9833 & 0.9694 & 0.9694 & 0.9575 \\
chb16 & Avg & 62.5 & 43.8 & 33.3 & 41.7 & \textbf{1.0000} & 0.8215 & 0.7237 & 0.7261 & \textbf{0.9833} & 0.9694 & 0.9694 & 0.9575 \\
\midrule
chb18 & 1 & 56.3 & 50.0 & 33.3 & 33.3 & 1.0000 & 0.9257 & 0.7846 & 0.7432 & 0.9820 & 0.9708 & 0.9733 & 0.9791 \\
chb18 & Avg & 56.3 & 50.0 & 33.3 & 33.3 & \textbf{1.0000} & 0.9257 & 0.7846 & 0.7432 & \textbf{0.9820} & 0.9708 & 0.9733 & 0.9791 \\
\midrule
chb21 & 1 & 139.6 & 139.6 & 125.0 & 139.6 & 1.0000 & 1.0000 & 0.9637 & 0.9985 & 0.9767 & 0.9767 & 0.9845 & 0.9805 \\
chb21 & Avg & 139.6 & 139.6 & 125.0 & 139.6 & \textbf{1.0000} & 1.0000 & 0.9637 & 0.9985 & 0.9767 & 0.9767 & \textbf{0.9845} & 0.9805 \\
\midrule
chb22 & 1 & 66.7 & 58.3 & 33.3 & 25.0 & 0.8477 & 1.0000 & 0.8121 & 
0.5709 & 0.9279 & 0.9291 & 0.9268 & 0.9163 \\
chb22 & 2 & 50.0 & 41.7 & 25.0 & 25.0 & 1.0000 & 0.8600 & 0.4047 & 0.5277 & 0.9429 & 0.9312 & 0.9302 & 0.9099 \\
chb22 & Avg & 58.3 & 50.0 & 29.2 & 25.0 & 0.9239 & \textbf{0.9300} & 0.6084 & 0.5493 & \textbf{0.9354} & 0.9302 & 0.9285 & 0.9131 \\
\midrule
Mean & - & 73.0 & 64.9 & 53.2 & 44.5 & 0.9878 & 0.9132 & 0.7881 & 0.7130 & 0.9518 & 0.9483 & 0.9506 & 0.9415 \\
$\pm$ Std &  & 35.8 & 38.9 & 40.0 & 38.8 & 0.0350 & 0.0975 & 0.1465 & 0.1466 & 0.0529 & 0.0519 & 0.0467 & 0.0491 \\
\bottomrule

\end{tabular}
\end{adjustbox}
\end{table*}

Entries in the CIOPR and F1-score columns highlighted in bold in Table~\ref{tab:allciopr} indicate the best performing preictal period length per patient for each metric. The 60-minute preictal length generally resulted in higher CIOPR scores, although the 45-minute one was optimal for some individuals, similar to findings in Section \ref{sub:f1}. The results can then be categorized into two types: a) concordant, where the same preictal length yielded the highest performance in both metrics and b) discordant, where CIOPR and F1-score values suggested different optimal preictal periods (OPP).\footnote{For case \textit{chb21}, the LOOCV approach (see Section \ref{sub:training}) resulted in training seizures with less than 45 minutes of preictal data when testing for the only eligible seizure (Seizure ID: 1) for CIOPR evaluation. Consequently, the results for the 60 and 45-minute preictal periods were identical and the 45-minute length was chosen since it provided a more accurate representation of the training instances.}

\begin{figure}[h]
        \centering
        \begin{subfigure}[b]{0.23\textwidth}
            \centering
            \includegraphics[width=\textwidth]{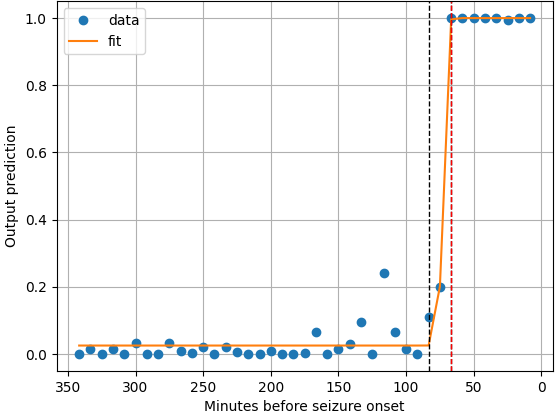}
            \caption[Network2]%
            {{\small 60 mins, CIOPR = 1.000, F1-Score = 0.9772}}    
            \label{fig:chb02_a}
        \end{subfigure}
        \hfill
        \begin{subfigure}[b]{0.23\textwidth}  
            \centering 
            \includegraphics[width=\textwidth]{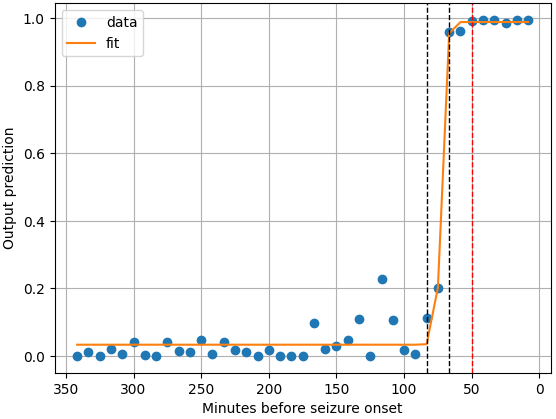}
            \caption[]%
            {{\small 45 mins, CIOPR = 0.8924, F1-Score = 0.9753}}    
            \label{fig:chb02_b}
        \end{subfigure}
        \vskip\baselineskip
        \begin{subfigure}[b]{0.23\textwidth}   
            \centering 
            \includegraphics[width=\textwidth]{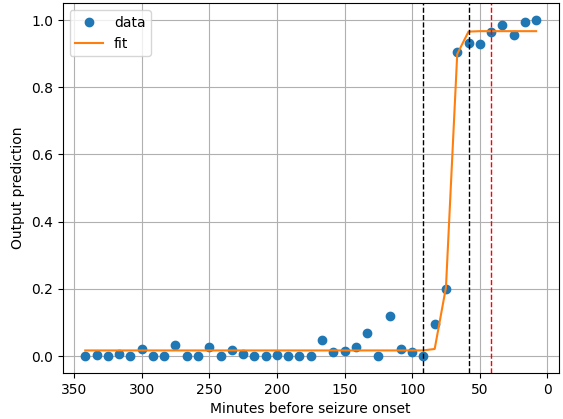}
            \caption[]%
            {{\small 30 mins, CIOPR = 0.8211, F1-Score = 0.9673}}    
            \label{fig:chb02_c}
        \end{subfigure}
        \hfill
        \begin{subfigure}[b]{0.23\textwidth}   
            \centering 
            \includegraphics[width=\textwidth]{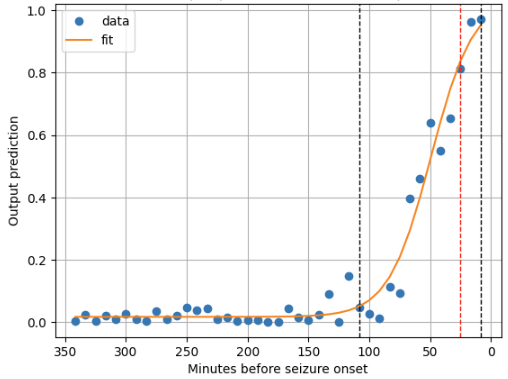}
            \caption[]%
            {{\small 15 mins, CIOPR = 0.6644, F1-Score = 0.9497}}    
            \label{fig:chb02_d}
        \end{subfigure}
        \caption[ chb02 ]
        {\small Output profiles of four patient-specific classifiers for case \textit{chb02} with 60, 45, 30, and 15 minutes preictal class definitions respectively. Input is 5.8 hours of continuous EEG recordings before the onset of seizure \#1 of \textit{chb02}. The red dotted line shows the seizure prediction convergence, and the black lines show the transition period boundaries. The horizontal axis represents minutes before seizure onset (seizure onset at utmost right), and the vertical axis is the classification output. Sub-captions show the preictal class definition, CIOPR value, and F1-score achieved.} 
        \label{fig:chb02_all}
    \end{figure}

Contrary to the F1-score results, the CIOPR values were more sensitive to varying preictal class definitions. In particular, a $25\%$ change was observed on average across different preictal lengths, reaching $>40\%$ in case \textit{chb14}. Cases \textit{chb02}, \textit{chb07}, \textit{chb14}, and \textit{chb16} all presented differences $\geq$10\% between the two best-performing models based on the CIOPR metric. Figure \ref{fig:chb02_all} depicts the output curves for case \textit{chb02} alongside the fitted sigmoidal curve. The CIOPR discrepancies can be attributed to the visually identifiable changes in the model's behavior, such as greater prediction horizon, shorter transition time, and reduced error. In this case, the results between the two metrics were concordant; increasing preictal class definition led to improved CIOPR and higher F1-scores. 

\begin{figure}[h]
        \centering
        \begin{subfigure}[b]{0.23\textwidth}
            \centering
            \includegraphics[width=\textwidth]{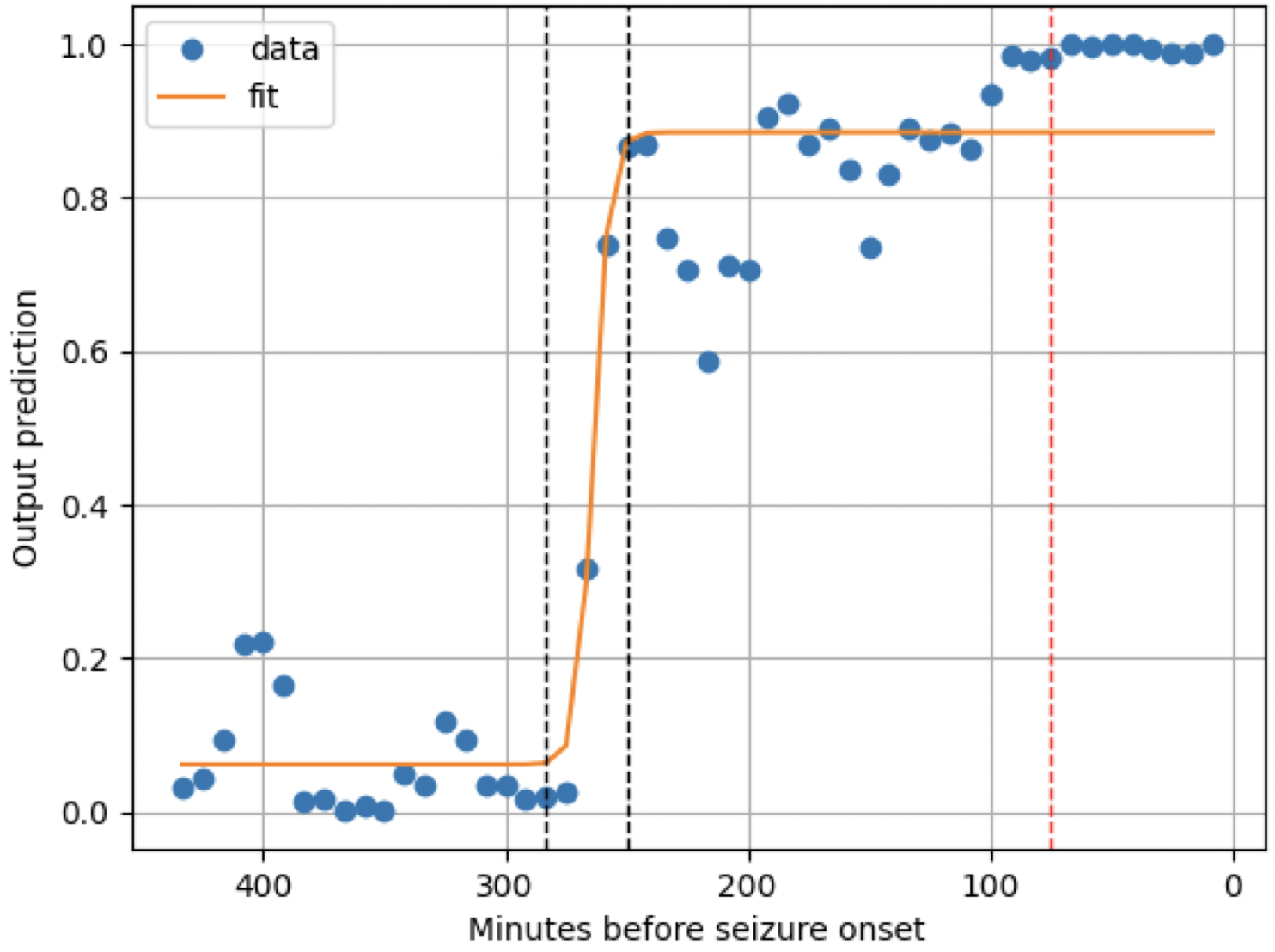}
            \caption[Network2]%
            {{\small 60 mins, CIOPR = 1.000, F1-Score = 0.9838}}    
            \label{fig:chb07_a}
        \end{subfigure}
        \hfill
        \begin{subfigure}[b]{0.23\textwidth}  
            \centering 
            \includegraphics[width=\textwidth]{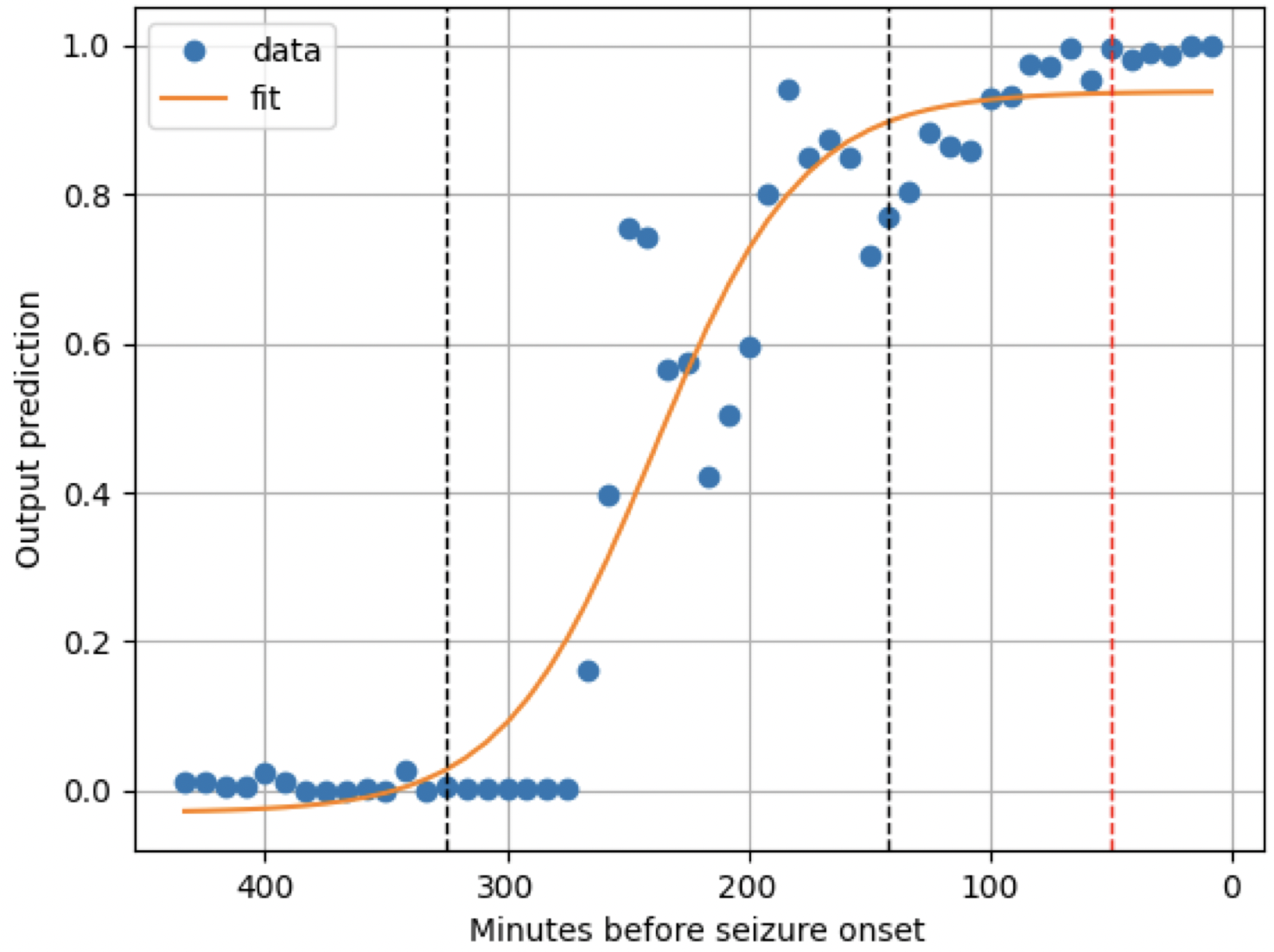}
            \caption[]%
            {{\small 45 mins, CIOPR = 0.8893, F1-Score = 0.9839}}    
            \label{fig:chb07_b}
        \end{subfigure}
        \vskip\baselineskip
        \begin{subfigure}[b]{0.23\textwidth}   
            \centering 
            \includegraphics[width=\textwidth]{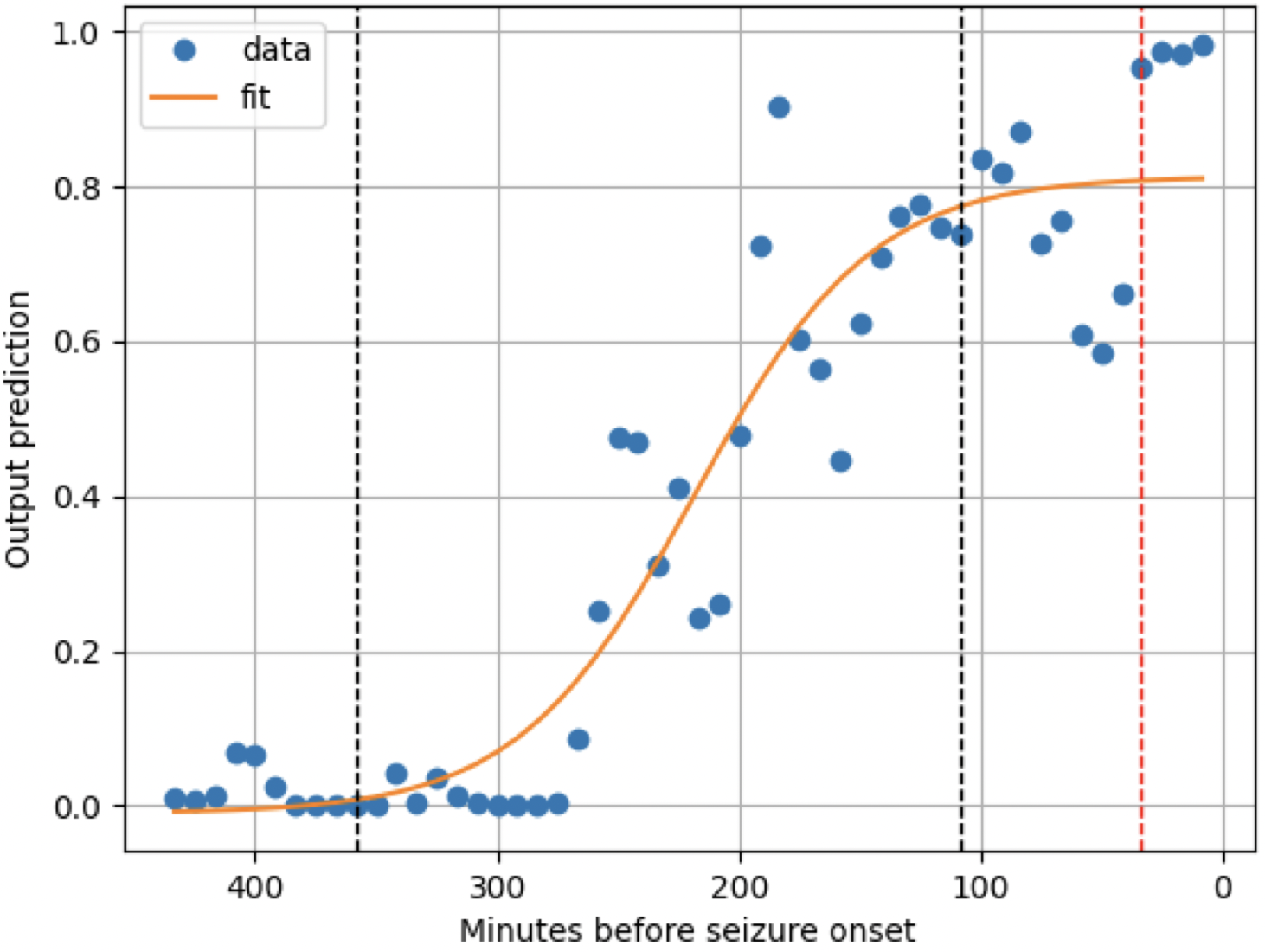}
            \caption[]%
            {{\small 30 mins, CIOPR = 0.7332, F1-Score = 0.9860}}    
            \label{fig:chb07_c}
        \end{subfigure}
        \hfill
        \begin{subfigure}[b]{0.23\textwidth}   
            \centering 
            \includegraphics[width=\textwidth]{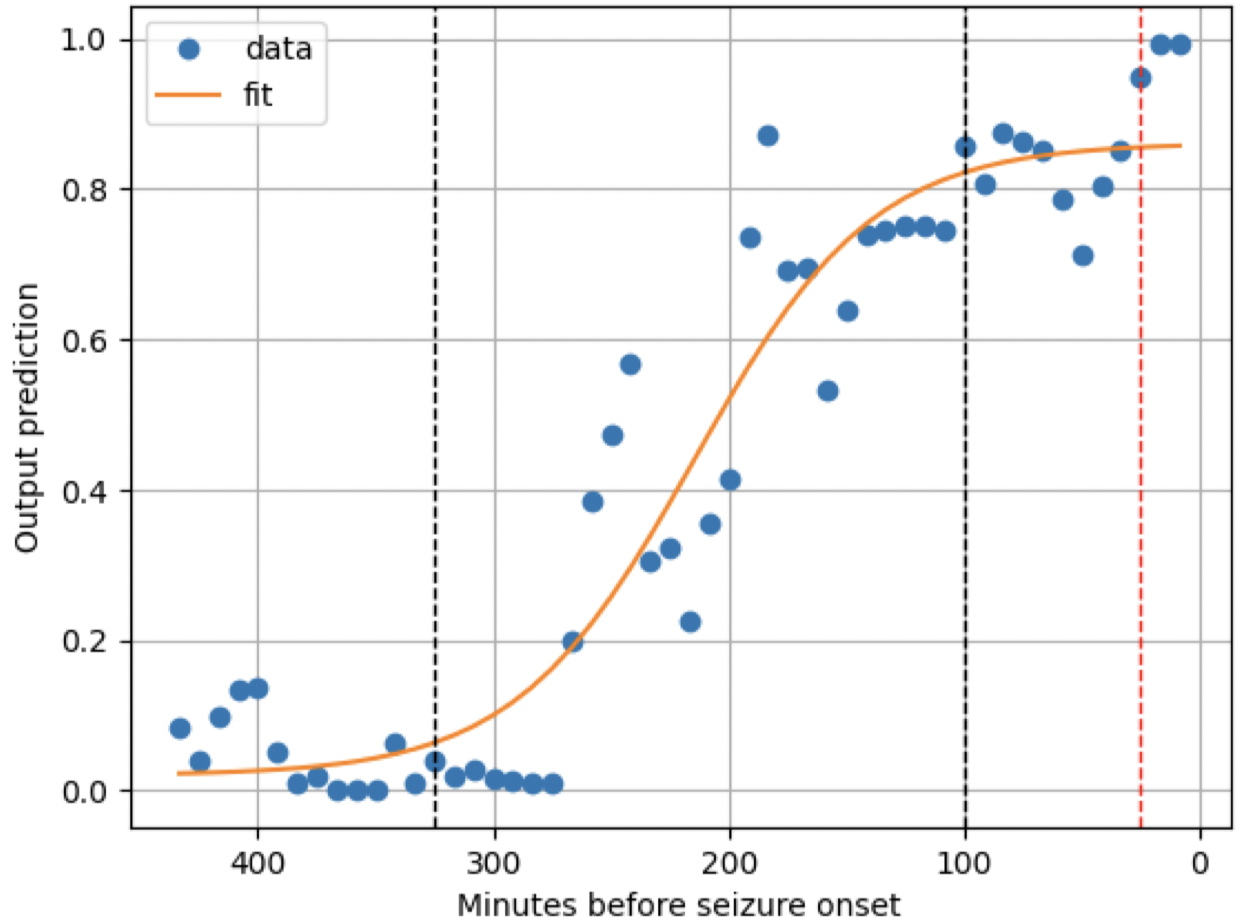}
            \caption[]%
            {{\small 15 mins, CIOPR = 0.7307, F1-Score = 0.9777}}    
            \label{fig:chb07_d}
        \end{subfigure}
        \caption[ chb07 ]
        {\small Output profiles of four patient-specific classifiers for case \textit{chb07} with 60, 45, 30, and 15 minutes preictal class definitions, respectively. Input is 7.0 hours of continuous EEG recordings before the onset of seizure \#1 of \textit{chb07}. The red dotted line shows the seizure prediction convergence, and the black lines show the transition period boundaries. The horizontal axis represents minutes before seizure onset (seizure onset at utmost right), and the vertical axis is the classification output. Sub-captions show the preictal class definition, CIOPR value, and F1-score achieved.} 
        \label{fig:chb07_all}
    \end{figure}

On the other hand, cases \textit{chb06}, \textit{chb07}, and \textit{chb14} demonstrated similar CIOPR variations ($\geq$10\%) across different preictal definitions, yet yielded discordant F1-score results. In particular, using the example of case \textit{chb07}, the differences in seizure prediction convergence and output stability are depicted in figure \ref{fig:chb07_all}. Longer preictal periods led to earlier prediction, shorter transition time and reduced output fluctuations, resulting in significantly greater CIOPR. Conversely, the F1-scores did not follow the same trend, underscoring the inadequacy of conventional metrics to fully capture the complex temporal behavior of seizure prediction models, as discussed in Section \ref{sub:introPreictal}.

As for the SPC results, they were generally aligned with CIOPR patterns, as anticipated by the design of the metric. However, cases \textit{chb01} and \textit{chb22}, which achieved the highest CIOPR values at a 45-minute preictal length, demonstrated a trade-off between minimized error, reduced transition time, and a slight decrease in prediction time, highlighting the integrative nature of the proposed metric. 

The classifiers' output convergence ranged from 217.6 to 16.7 minutes before the seizure onset, occurring on average earlier at longer preictal state definitions. Seizure 2, case \textit{chb04} was the only exception, where the classifier converged 9.3 minutes earlier when the preictal length was reduced to 45 minutes. Nevertheless, the greatest variability was observed across patients rather than preictal definitions, with a standard deviation of $\pm$ 35.8 minutes for the 60-minute preictal duration, accounting for approximately $50\%$ of the mean value.

\subsection{OPP selection and overall performance}
\label{sub:overall}

For seizures eligible for CIOPR assessment, the OPP was determined based on the highest CIOPR values from Table \ref{tab:allciopr}, while for the rest, it was based on the highest F1-score from Table \ref{tab:test1_table}, as outlined in Section \ref{sub:testing}. Overall results are summarized in Table \ref{tab:combined_table}, which includes the OPP for each patient and corresponding performance metrics. $SEN$, $SPE$, $FAR$, $ACC$, $AUC$, and F1-score were computed for the selected OPP using all testing seizures as in Table \ref{tab:test1_table}, while SPC used only the ones subject to CIOPR testing. The criterion used to select the OPP (CIOPR or F1-score) is also displayed. On average, the subject-specific models achieved a sensitivity of $99.31\%$, specificity of $95.34\%$, classification accuracy of $97.32\%$, and F1-score of $97.46\%$. The classifiers' output converged at 76.8 minutes before seizure onset, with a standard deviation of 36.8 minutes, reflecting anticipated cross-patient heterogeneities.

\begin{table*}[htbp]
\centering
\caption{Overall Results for Selected Preictal Definition per Patient}
\label{tab:combined_table}
\begin{adjustbox}{max width=\textwidth}
\begin{tabular}{@{}lcccccccccccccccc@{}}
\toprule
Case & Sensitivity (\%) & Specificity (\%) & FAR ($h^{-1}$) & Accuracy (\%) & F1-score (\%) & AUC (\%) & SPC (mins) & OPP (mins) & Criterion \\ 
\midrule
chb01 & 100.0 & 99.85 & 1.06 & 99.93 & 99.93 & 99.99 & 61.8 & 45 & CIOPR \\
chb02 & 99.76 & 96.32 & 26.5 & 98.04 & 98.08 & 99.86 & 66.7 & 60 & CIOPR \\
chb03 & 99.87 & 97.56 & 17.6 & 98.71 & 98.73 & 99.93 & N/A & 30 & F1-Score \\
chb04 & 99.84 & 99.19 & 5.83 & 99.52 & 99.40 & 99.99 & 175.5 & 45 & CIOPR \\
chb05 & 99.62 & 96.66 & 24.1 & 98.14 & 98.16 & 99.87 & 54.9 & 60 & CIOPR \\
chb06 & 94.76 & 73.54 & 190 & 84.15 & 85.71 & 92.90 & 52.9 & 60 & CIOPR \\
chb07 & 99.54 & 97.74 & 16.2 & 98.64 & 98.66 & 99.72 & 61.5 & 60 & CIOPR \\
chb09 & 99.72 & 98.85 & 8.30 & 99.28 & 99.29 & 99.98 & 93.5 & 60 & CIOPR \\
chb10 & 98.40 & 92.84 & 51.6 & 95.62 & 95.76 & 99.54 & 68.8 & 60 & CIOPR \\
chb11 & 99.86 & 99.81 & 1.37 & 99.83 & 99.83 & 99.99 & N/A & 45 & F1-Score \\
chb14 & 97.85 & 86.81 & 95.0 & 92.33 & 92.73 & 97.68 & 52.1 & 60 & CIOPR \\
chb16 & 99.64 & 97.00 & 21.6 & 98.32 & 98.33 & 99.82 & 62.5 & 60 & CIOPR \\
chb17 & 99.45 & 97.57 & 17.5 & 98.51 & 98.51 & 99.82 & N/A & 60 & F1-Score \\
chb18 & 99.59 & 96.40 & 25.9 & 98.00 & 98.02 & 99.43 & 56.3 & 60 & CIOPR \\
chb19 & 100.0 & 99.64 & 2.60 & 99.82 & 99.82 & 100.0 & N/A & 30 & F1-Score \\
chb20 & 100.0 & 98.84 & 8.37 & 99.39 & 99.42 & 99.99 & N/A & 15 & F1-Score \\
chb21 & 99.74 & 97.03 & 21.4 & 98.37 & 98.41 & 99.82 & 139.6 & 45 & CIOPR \\
chb22 & 99.19 & 86.38 & 98.1 & 92.79 & 93.23 & 99.38 & 50.0 & 45 & CIOPR \\
chb23 & 100.0 & 99.36 & 4.60 & 99.68 & 99.68 & 99.99 & N/A & 60 & F1-Score \\
\midrule
Mean & 99.31 & 95.34 & 33.6 & 97.32 & 97.46 & 99.35 & 76.8 &  &  \\
$\pm$ Std & 1.20 & 6.38 & 46.0 & 3.76 & 3.41 & 1.61 & 36.8 &  &  \\
\bottomrule
\end{tabular}
\end{adjustbox}
\end{table*}

\subsection{Statistical analysis results}
\label{sub:statistical_results}
The null hypothesis formulated in Section \ref{sub:statistical_analysis} was rejected with Bonferroni-corrected p-values $<0.05$ for all metrics apart from $ND_{err}$ (p-value $=0.388$), $TP$ (p-value $=0.135$), and sensitivity (p-value $=0.161$). The comparison between the 60 and 15-minute preictal periods showed significant differences in 8 of the 11 metrics assessed, followed by the 45 to 15, and 60 to 30-minute comparisons. Conversely, the smallest number of significant differences were noted between adjacent preictal definitions (15 minutes apart), particularly for the 30 to 15-minute and 60 to 45-minute pairs. CIOPR and SPC metrics accounted for the highest number of significant comparisons (4 out of 6), followed by specificity (3 out of 6), accuracy (2 out of 6), and lastly by the F1-score (1 out of 6). Detailed p-values across all the hypothesis tests performed can be found in the Supplementary Material, Table 5.

\section{Discussion}
\label{sec:discussion}
We trained subject-specific classifiers for 19 pediatric patients of the CHB-MIT dataset to predict epileptic seizures using raw scalp EEG signals. The proposed CNN-Transformer deep learning architecture achieved a balanced accuracy of $97.32\%$, enabling confident seizure predictions up to 76.8 minutes prior to seizure onset on average. By introducing the CIOPR metric, we illustrated the significant impact of preictal class definition on model behavior and demonstrated how these variations can be quantified to determine the optimal preictal duration to maximize the model's usability for each individual patient.

\subsection{Classification performance}
\label{sub:discussion-performance}
The proposed architecture demonstrated consistent performance across multiple subjects and preictal lengths with particularly high sensitivity ($>99\%$). We used the balanced accuracy metric (average of sensitivity and specificity) for comparison with state-of-the-art literature \cite{LSTM, AttentionCNNGRU, ryu_2021, jemal_2022, baghdadi_2020, affes_2022, ma_2024, qi_2024, yang_2021, rasheed_2021, jana_2021, ibrahim_2022} to account for variations in the preictal-interictal data ratio across studies. Our model achieved higher balanced accuracy than the aforementioned studies, by an average of $7\%$. While the authors in \cite{LSTM} reported the highest accuracy value, the absence of validation using LOOCV undermines the reported performance. To the best of our knowledge, Daoud \textit{et al.} \cite{Daoud} have attained the highest LOOCV-validated accuracy of $99.66\%$. Detailed patient-level performance comparison with existing literature can be found in the Supplementary Material, Table 6. False alarms averaged 33.6 per hour on a segment-wise basis, since they were computed on the raw model output. However, the average specificity exceeded $95\%$, suggesting that the introduction of a post-processing scheme would significantly reduce FAR ($h^{-1}$) in a clinical environment.  

Although extended preictal periods could be expected to increase the likelihood of mislabeling interictal data, and consequently the number of false positives, the results of our study indicate the opposite. Varying preictal definitions minimally impacted model sensitivity as shown in Section \ref{sub:statistical_results}, whereas specificity improved with longer preictal periods. This finding aligns with the results reported in literature \cite{LSTM}. Furthermore, at the 0.5 decision threshold, specificity was consistently lower than sensitivity, a pattern appearing also in other studies \cite{jemal_2022, affes_2022, qi_2024, rasheed_2021, jana_2021}. 

Higher sensitivity scores and consequent invariability to changing preictal state definitions may reflect one of the major limitations in ongoing epileptic seizure prediction research: insufficient data volume. Data augmentation in the preictal state, along with a lack of external validation, could lead to model overfitting, where classifiers memorize rather than generalize training data \cite{DLEpilepsy}. Although the LOOCV approach prevents the model from encountering test seizures during training, the uniform data acquisition setting and the close temporal proximity of recordings might still result in high similarity between training and testing sets \cite{detection_review}. Conversely, interictal data, recorded throughout the whole day, exhibit greater variability from circadian rhythms and additional EEG patterns \cite{circadian}, complicating their classification. This greater variability might explain the consistently lower specificity and its significant reduction with shorter preictal durations, as these entail fewer training samples. 

Some cases in our study had significantly lower performance compared to others. For instance, case \textit{chb06} exhibited a sensitivity of $94.76\%$, and a specificity of $73.54\%$, $>20\%$ below the average. Cases \textit{chb14} and \textit{chb22} also showed notably low specificity scores. Comparison with literature suggests that cases \textit{chb06} and \textit{chb14} exhibited on average the lowest balanced accuracy across state-of-the-art studies. The same pattern was observed when comparing the reported F1-scores in \cite{affes_2022, qi_2024, yan_2023, ibrahim_2022, ryu_2021, assali_2023, jemal_2022}. The limited number of patients and the absence of detailed pathological data in the CHB-MIT dataset prevent further exploration of the effect of various epilepsy sub-types on seizure prediction performance. However, concordance with the literature results indicates that the aforementioned limitations are broadly applicable to the field and could inform future research directions.

\subsection{CIOPR alongside conventional metrics}
\label{sub:discussion-ciopr}
The newly introduced CIOPR metric was used to comprehensively assess model performance and compare the effect of different preictal state definitions. Comparison with the F1-score results indicates that CIOPR is considerably more sensitive to varying preictal lengths, showing an average change of $9.2\%$ per 15-minute decrease, compared to a $0.3\%$ average change for the F1-score. This is also confirmed by the statistical analysis in Section \ref{sub:statistical_results}, where the CIOPR distributions varied significantly across four preictal length comparisons, compared to only one for the F1-score. Furthermore, in 7 out of the 13 cases that underwent CIOPR testing, there was a discordance in the best-performing preictal definition between the two metrics. This discrepancy occurred only in cases where F1-score variations were less than $1\%$ between the two best-performing preictal definitions.

\begin{figure*}[htbp]
    \centering
    \begin{subfigure}[b]{0.24\textwidth}
        \centering
        \includegraphics[width=\textwidth, height=5cm]{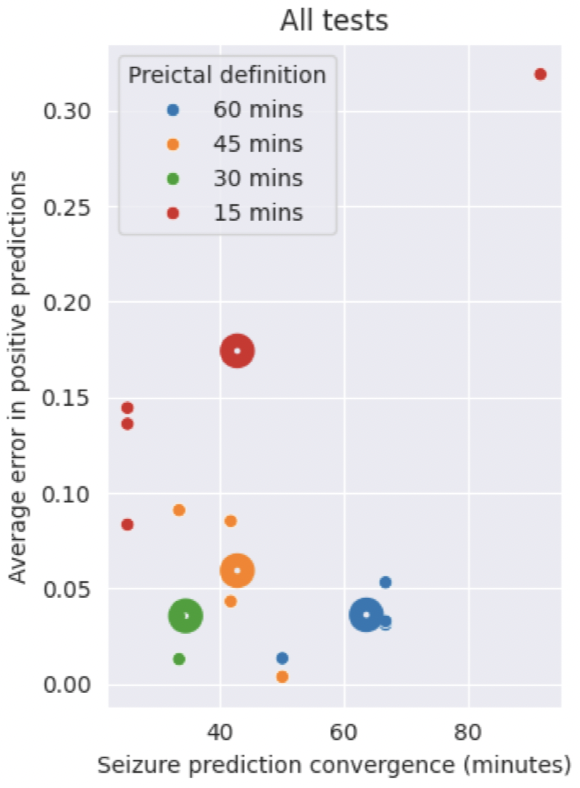}
        \caption{\small Average error in positive predictions versus SPC.}
        \label{fig:subfig1}
    \end{subfigure}
    \hfill
    \begin{subfigure}[b]{0.24\textwidth}
        \centering
        \includegraphics[width=\textwidth, height=5cm]{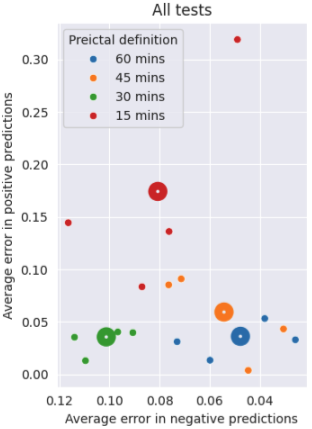}
        \caption{\small Average error in positive versus negative predictions.}
        \label{fig:subfig2}
    \end{subfigure}
    \hfill
    \begin{subfigure}[b]{0.24\textwidth}
        \centering
        \includegraphics[width=\textwidth, height=5cm]{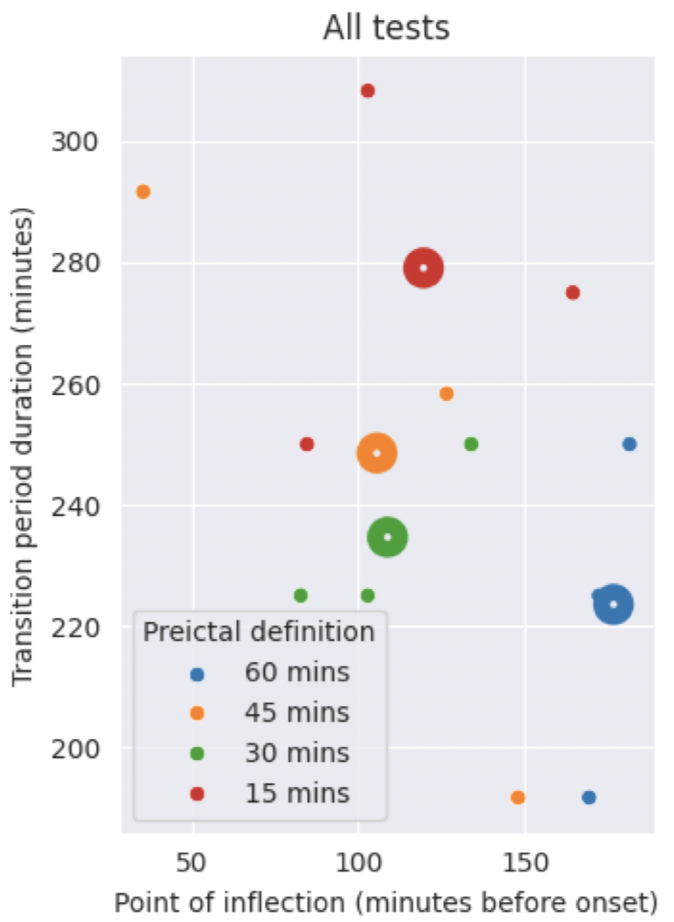}
        \caption{\small Transition period duration versus point of inflection.}
        \label{fig:subfig3}
    \end{subfigure}
    \hfill
    \begin{subfigure}[b]{0.24\textwidth}
        \centering
        \includegraphics[width=\textwidth, height=5cm]{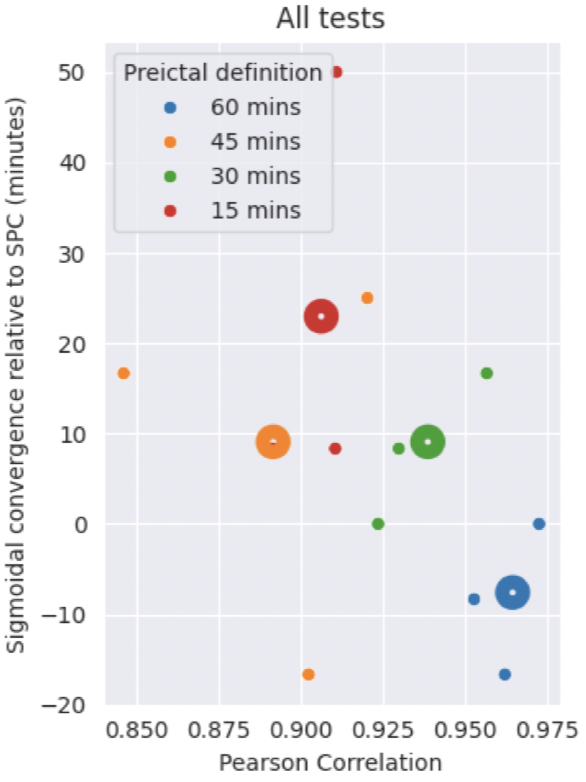}
        \caption{\small Convergence points versus Pearson correlation.}
        \label{fig:subfig4}
    \end{subfigure}
    \caption{\small Performance measures used for CIOPR calculation of seizure \#1, case \textit{chb16} across different preictal state definitions. The best performance corresponds to the bottom-right corner. Sub-figure \ref{fig:subfig4} assesses the quality of the sigmoidal fitting, where the difference between the convergence points of the fitted curve and the model's output (SPC) are shown in the vertical axis. Dots in the plots represent different training runs (tests) for a given preictal, while larger circles show the mean values.}
    \label{fig:interpretability}
\end{figure*}

While the F1-score and other conventional metrics remain useful and can guide design choices, comparing high-performing classifiers necessitates more sophisticated and sensitive measures to capture subtle behavioral differences. Figure \ref{fig:interpretability} illustrates how the individual measures used to calculate the CIOPR allow a comprehensive assessment of the model's behavior for case \textit{chb16}. The performance advantages of a 60-minute preictal definition -- early prediction time, minimized positive and negative errors, short transition period, and alignment with the fitted sigmoidal curve representing ideal behavior -- are visualized in an interpretable manner that conventional metrics lack.

\subsection{Preictal length definition and model behavior}
\label{sub:discussion-ciopr2}
Observing the CIOPR scores and individual measures (detailed in the Supplementary Material, Table 4) across all patients enabled us to draw general conclusions about the model's behavior under different preictal state definitions. As discussed in Section \ref{sub:discussion-performance}, the rate of false positives increased with shorter preictal lengths, while sensitivity remained relatively constant. Conversely, CIOPR results showed that the error during negative predictions ($ND_{err}$, Section \ref{sub:methodsCIOPR}) increased with longer preictal definitions. This can be explained by $ND_{err}$ being measured up to 10 hours before onset, unlike specificity, which was calculated using all interictal data. However, due to the large percent variations in $ND_{err}$ and the limited number of testing seizures, the effect of preictal definitions in the $ND_{err}$ distributions was deemed not statistically significant. 

Models trained with longer preictal periods could detect preictal-like dynamics many hours before onset, leading to an earlier gradual increase in positive predictions: the point of inflection in the fitted sigmoidal curve occurred on average 143.1 minutes before onset for the 60-minute definition, compared to 125.3 minutes for the 15-minute definition. Statistical analysis showed that reducing the preictal length from 60 to 15 and from 60 to 30 minutes led to significantly different distributions for the point of inflection. Therefore, these early positive predictions also reflected by the increased $ND_{err}$ could be related to the impending seizure. However, they could cause patient distress and be considered undesired in a clinical setting. Similarly, longer preictal definitions led to earlier convergence of the classifier output, averaging 73 minutes before onset. In combination with lower $SPC_{err}$ values and higher correlation coefficients with the fitted curve, increasing the preictal length appeared to lead to more accurate predictions in the preictal state with reduced output fluctuations.  

Although the transition from interictal to preictal predictions occurred faster in the 45-minute model compared to the 60-minute model, this difference was not statistically significant. Longer preictal periods allow classifiers to learn spatiotemporal EEG features more distantly from the seizure onset, enhancing state transition clarity. However, the slightly faster transition observed in the 45-minute model suggests that features learned more than 45 minutes before onset might overlap with interictal features and lose relevance for prediction. Conversely, features learned within the last 30 minutes before onset achieve high prediction accuracy but may be insufficient for early identification of an impending seizure. This could lead to increased uncertainty during the transition period and potentially contradictory outputs in a clinical implementation setting. This is further highlighted by the fact that among adjacent preictal definitions, the 45 to 30-minute comparison yielded the highest number of significant differences across all metrics. 

Overall, a 60-minute preictal duration achieved the highest CIOPR scores, primarily due to the tuning of the algorithm to positively reward early prediction times. However, the CIOPR metric's multi-faceted design also emphasizes reducing the prediction error, leading to an OPP of 45 minutes for cases \textit{chb01} and \textit{chb22}, despite having shorter prediction times. At the OPP, seizures were accurately predicted on a patient-wise average of 76.8 minutes before the onset, which to the best of our knowledge, is the earliest reported in the literature (detailed comparison in the Supplementary Material, Table 7).

\subsection{Cross- and intra-patient heterogeneities}
\label{sub:discussion-ciopr3}
The CIOPR-related measures and their sensitivity to different preictal period lengths varied across testing seizures. This observation underscores the major challenge in epileptic seizure prediction research: inter-seizure heterogeneities, both across and within patients. These differences caused the optimal preictal period (OPP) to fluctuate between 60 and 45 minutes across patients, and in some cases (e.g., \textit{chb01}, \textit{chb22}) within seizures of the same patient. Cross-patient variations can be addressed by designing patient-specific models. Intra-patient variations can be managed by choosing the best average across seizures, as in our study, or selecting a different preictal length definition for each training seizure, as suggested by \cite{OPP}. 

These approaches, though effective at improving the classification of interictal and preictal segments, cannot eliminate inconsistent values across measures comprising CIOPR, due to the unique electrophysiological signature of epileptic seizures. Among these measures, prediction time is of utmost importance as it directly influences alarm generation and provides a window for potential intervention.
Keeping two variables among \textit{Patient ID, Seizure ID, Preictal Length} fixed, and computing the mean standard deviation on the third showed that variations in prediction time could be primarily attributed to cross-patient heterogeneities ($\pm$38.3 minutes), followed by intra-patient heterogeneities ($\pm$17.7 minutes), and lastly, by the preictal period lengths used for training ($\pm$13.2 minutes).

The clinical usefulness of a seizure prediction system would rely on meeting the patients' preferences in prediction time and sensitivity-specificity trade-off \cite{wristband_sensor}. Researchers in \cite{first_in_man_study} highlighted that patient satisfaction was maximized with short prediction times, allowing effective lifestyle changes. Long prediction times on the other hand could be more useful in a closed-loop system, allowing room for drug-delivery or stimulation \cite{prediction_review}. However, the observed cross-patient variations suggest that prediction time is not merely a matter of preference but is intrinsically linked to the preictal characteristics of each patient. Additionally, intra-patient variations, although less pronounced, could significantly impact the effectiveness of seizure predictions. For instance, in case \textit{chb09}, the $SPC$ ranged from 63.9 to 125 minutes. 

%Commenting out the following sentences since seizure prediction is still at an early stage.
\iffalse
Such inconsistencies can be detrimental as they may cause patients to misjudge the importance of warnings and undermine their ability to take appropriate actions. Furthermore, uncertainty following alarms can increase confusion and potentially lead to adverse effects on patient well-being.
\fi

A reliable system should predict seizures earlier than a minimum seizure prediction horizon (SPH) with fluctuations not exceeding the seizure occurrence period (SOP), such that prediction times range within the $[SPH, SPH+SOP]$ interval \cite{sph_sop}. SPH should be large enough to allow clinical intervention (e.g., 30 minutes in \cite{sph_sop}), while SOP could be limited to the duration of the treatment; e.g., 30 minutes for some anti-epileptic drugs \cite{sph_sop}. Cross-patient heterogeneities could enable setting a realistic SPH interval, informed by both the patient's preferences and the electrophysiological signature of the preictal state. Case \textit{chb04} for instance exhibited $SPC$ values of 133.3 and 217.5 minutes before the onset. A 30-minute SPH selection would hence prove non-practical for this case, as it would require a SOP of at least 3 hours, leading to increased uncertainty for the patient. Similarly, understanding intra-patient heterogeneity patterns could allow providing realistic SOP intervals for each individual. For instance, at the OPP, case \textit{chb07} experienced considerably less variations in prediction time ($\pm$4.7 minutes) than case \textit{chb09} ($\pm$30.6 minutes).

\subsection{Limitations and future work}
\label{sub:limitations}
Besides the aforementioned risk of overfitting, performance results could be affected by data leakage, a prevalent issue in machine learning-related research \cite{leakage}. In particular, the LOOCV approach might introduce temporal leakage, due to the inclusion of ``future'' data (i.e., seizures) in the training process. For example, a model trained on seizures \textit{No. 1, 3, 4, 5} and tested on seizure \textit{No. 2} had access to information that occurred both before and after the test seizure, potentially introducing temporal dependencies between the training and testing sets. This information would not be available in a realistic implementation setting and could inflate the observed performance. 

Furthermore, the classifiers' output behavior was not uniform across all subjects and seizures, introducing limitations in the generalizability of the \textit{CIOPR} metric. Convergence did not always occur right before seizure onset (Figure \ref{fig:corr_b}), while in several cases, persistent output fluctuations complicated the identification of the $SPC$ start (Figure \ref{fig:corr_d}). Additionally, the $TP$ was directly computed from the fitted curve, making it dependent on the quality of the fitting. Figure \ref{fig:chb07_all} highlights the issue, where small variations in model behavior caused large differences in the stretch of the sigmoidal curve and consequently in the calculated $TP$. Lastly, the penalty introduced due to long transition periods was dependent on the interictal duration, $ND$, inhibiting direct comparisons across different seizures. 

Future efforts should prioritize improving the quality of EEG datasets. Extended recording times and longitudinal acquisitions are necessary to assess algorithms' generalizability over time. Such a dataset could alleviate the need for the LOOCV approach and consequently mitigate data leakage issues. Reporting pathological information, such as the location of the seizure onset zone, clinical semiology, and epilepsy sub-type, would be crucial in drawing clinically relevant conclusions, including which seizure sub-types could be more challenging to predict. Additionally, it could enable an understanding of preictal state-related heterogeneities across seizures and patients. Unraveling those patterns could enable tailored identification of SPH and SOP intervals, that could be dynamically adjusted during the course of the day, depending on the individual and epilepsy pathology. This would ultimately lead to providing context-aware warnings and improving the clinical reliability of automated seizure prediction models.

The proposed solution would also benefit from developing a channel-selection algorithm at the input stage, as suggested by authors in \cite{affes_2022}. Furthermore, implementing a post-processing scheme to filter out incorrect predictions and generate confident alarms would be useful.  Clinically relevant metrics such as FAR ($h^{-1}$) should then be computed on the post-processed output and reported for a predefined SPH and SOP. While the CIOPR algorithm prioritizes prolonged prediction times, future improvements should tune it to emphasize convergence within the preferred $[SPH, SPH+SOP]$ range. Consistent EEG datasets will enable benchmarking on achieved CIOPC values by fixing the input signal duration and the contribution of each term (e.g., transition period).

\section{Conclusions}
\label{sec:conclusion}
We introduced a CNN-Transformer model for predicting epileptic seizures using scalp EEG signals and proposed the novel Continuous Input-Output Performance Ratio (CIOPR) metric.Using the CIOPR metric we established that varying preictal period lengths result in statistically significant differences in model behavior. Overall, increasing the preictal length led to earlier prediction times, sharper interictal-preictal transitions, and reduced output fluctuations in the preictal state, at the expense of an increased volume of positive predictions several hours before the onset. This finding highlights the need for careful selection of the OPP depending on the desired usability. 

To the best of our knowledge, this is the first study that uses a fitting curve to model the output of a seizure prediction classifier. The newly introduced CIOPR metric provides interpretability on model performance that conventional metrics lack, by outlining how the distribution of positive predictions varies over time. With the increasing complexity and classification capabilities of deep learning architectures, more intuitive and sophisticated measures are required to capture subtle performance differences. Integrative measures like CIOPR, apart from being useful in selecting the most suitable deep learning model, they can provide meaningful clinical insights and shed light on future research directions.

% or


\section*{Acknowledgment}

The authors would like to acknowledge the Epilepsy Research UK Doctoral Training Centre for providing the scholarship to BC.

\ifCLASSOPTIONcaptionsoff
  \newpage
\fi



\bibliographystyle{IEEEtran}
\bibliography{bibliography}


\begin{comment}
\begin{IEEEbiography}{Michael Shell}
Biography text here.
\end{IEEEbiography}

\begin{IEEEbiographynophoto}{John Doe}
Biography text here.
\end{IEEEbiographynophoto}

%\newpage

\begin{IEEEbiographynophoto}{Jane Doe}
Biography text here.
\end{IEEEbiographynophoto}


%\vfill


\includepa

\end{comment}



\section*{Supplementary material}

# Supplementary Materials for "Preictal Period Optimization for Deep Learning-Based Epileptic Seizure Prediction"

Petros Koutsouvelis, Bartlomiej Chybowski, Alfredo Gonzalez-Sulser, Shima Abdullateef, Javier Escudero

## 1. Patient Information & Data Selection:

Table 1: Information for each case of the CHB-MIT dataset, including number of eligible seizures (eligibility criteria defined in the manuscript) and seizures applicable for CIOPR testing.

| Case | Sex | Age | All Seizures | Eligible | CIOPR Testing |
|------|-----|-----|--------------|----------|---------------|
| chb01 | F | 11 | 7 | 5 | 3 |
| chb02 | M | 11 | 3 | 3 | 1 |
| chb03 | F | 14 | 7 | 6 | 0 |
| chb04 | M | 22 | 4 | 3 | 2 |
| chb05 | F | 7 | 5 | 4 | 3 |
| chb06 | F | 1.5 | 10 | 10 | 6 |
| chb07 | F | 14.5 | 3 | 3 | 3 |
| chb08 | M | 3.5 | 5 | - | - |
| chb09 | F | 10 | 4 | 4 | 3 |
| chb10 | M | 3 | 7 | 7 | 1 |
| chb11 | F | 12 | 3 | 3 | 0 |
| chb12 | F | 2 | 40 | - | - |
| chb13 | F | 3 | 12 | - | - |
| chb14 | F | 9 | 8 | 5 | 2 |
| chb15 | M | 16 | 20 | - | - |
| chb16 | F | 7 | 10 | 3 | 1 |
| chb17 | F | 12 | 3 | 3 | 0 |
| chb18 | F | 18 | 6 | 5 | 1 |
| chb19 | F | 19 | 3 | 3 | 0 |
| chb20 | F | 6 | 8 | 4 | 0 |
| chb21 | F | 13 | 4 | 4 | 1 |
| chb22 | F | 9 | 3 | 3 | 2 |
| chb23 | F | 6 | 7 | 5 | 0 |
| chb24 | - | - | 16 | - | - |
| Total | - | - | 198 | 83 | 29 |
| Average | - | - | 8.3 | 4.4 | 1.5 |

Table 2: Patient ID, Seizure ID, and corresponding file name in the CHB-MIT dataset for each eligible seizure, including applicability for CIOPR testing (continued on next page).

| Patient ID | Seizure ID | File Name | CIOPR [Y/N] |
|---|---|---|---|
| chb01 | 1 | chb01_03.edf | Y |
| chb01 | 2 | chb01_15.edf | Y |
| chb01 | 3 | chb01_18.edf | N |
| chb01 | 4 | chb01_21.edf | N |
| chb01 | 5 | chb01_26.edf | Y |
| chb02 | 1 | chb02_16.edf | Y |
| chb02 | 2 | chb02_16+.edf | N |
| chb02 | 3 | chb02_19.edf | N |
| chb03 | 1 | chb03_01.edf | N |
| chb03 | 2 | chb03_02.edf | N |
| chb03 | 3 | chb03_04.edf | N |
| chb03 | 4 | chb03_34.edf | N |
| chb03 | 5 | chb03_35.edf | N |
| chb03 | 6 | chb03_36.edf | N |
| chb04 | 1 | chb04_05.edf | Y |
| chb04 | 2 | chb04_08.edf | Y |
| chb04 | 3 | chb04_28.edf | N |
| chb05 | 1 | chb05_06.edf | Y |
| chb05 | 2 | chb05_13.edf | Y |
| chb05 | 3 | chb05_16.edf | N |
| chb05 | 4 | chb05_22.edf | Y |
| chb06 | 1 | chb06_01.edf | N |
| chb06 | 2 | chb06_01.edf | N |
| chb06 | 3 | chb06_01.edf | N |
| chb06 | 4 | chb06_04.edf | Y |
| chb06 | 5 | chb06_04.edf | N |
| chb06 | 6 | chb06_09.edf | Y |
| chb06 | 7 | chb06_10.edf | Y |
| chb06 | 8 | chb06_13.edf | Y |
| chb06 | 9 | chb06_18.edf | Y |
| chb06 | 10 | chb06_24.edf | Y |
| chb07 | 1 | chb07_12.edf | Y |
| chb07 | 2 | chb07_13.edf | Y |
| chb07 | 3 | chb07_19.edf | Y |
| chb09 | 1 | chb09_06.edf | Y |
| chb09 | 2 | chb09_08.edf | Y |
| chb09 | 3 | chb09_08.edf | N |
| chb09 | 4 | chb09_19.edf | Y |

| Patient ID | Seizure ID | File Name | CIOPR [Y/N] |
|---|---|---|---|
| chb10 | 1 | chb10_12.edf | N |
| chb10 | 2 | chb10_20.edf | Y |
| chb10 | 3 | chb10_27.edf | N |
| chb10 | 4 | chb10_30.edf | N |
| chb10 | 5 | chb10_31.edf | N |
| chb10 | 6 | chb10_38.edf | N |
| chb10 | 7 | chb10_89.edf | N |
| chb11 | 1 | chb11_82.edf | N |
| chb11 | 2 | chb11_92.edf | N |
| chb11 | 3 | chb11_99.edf | N |
| chb14 | 1 | chb14_03.edf | Y |
| chb14 | 2 | chb14_06.edf | N |
| chb14 | 3 | chb14_11.edf | N |
| chb14 | 4 | chb14_17.edf | N |
| chb14 | 5 | chb14_27.edf | Y |
| chb16 | 1 | chb16_10.edf | Y |
| chb16 | 2 | chb16_14.edf | N |
| chb16 | 3 | chb16_16.edf | N |
| chb17 | 1 | chb17a_03.edf | N |
| chb17 | 2 | chb17a_04.edf | N |
| chb17 | 3 | chb17b_63.edf | N |
| chb18 | 1 | chb18_29.edf | Y |
| chb18 | 2 | chb18_31.edf | N |
| chb18 | 3 | chb18_32.edf | N |
| chb18 | 4 | chb18_35.edf | N |
| chb18 | 5 | chb18_36.edf | N |
| chb19 | 1 | chb19_28.edf | N |
| chb19 | 2 | chb19_29.edf | N |
| chb19 | 3 | chb19_30.edf | N |
| chb20 | 1 | chb20_12.edf | N |
| chb20 | 2 | chb20_13.edf | N |
| chb20 | 3 | chb20_16.edf | N |
| chb20 | 4 | chb20_68.edf | N |
| chb21 | 1 | chb21_19.edf | Y |
| chb21 | 2 | chb21_20.edf | N |
| chb21 | 3 | chb21_21.edf | N |
| chb21 | 4 | chb21_22.edf | N |
| chb22 | 1 | chb22_20.edf | Y |
| chb22 | 2 | chb22_25.edf | Y |
| chb22 | 3 | chb22_38.edf | N |
| chb23 | 1 | chb23_06.edf | N |
| chb23 | 2 | chb23_08.edf | N |
| chb23 | 3 | chb23_08.edf | N |
| chb23 | 4 | chb23_09.edf | N |
| chb23 | 5 | chb23_09.edf | N |

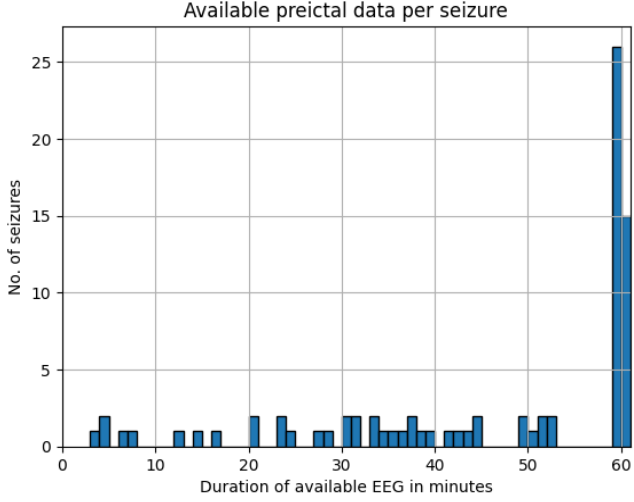

Figure 1: Distribution of available preictal EEG data duration for each eligible seizure. Maximum duration of 60 minutes, according to the data selection criteria stated in the manuscript.

## 2. Deep Learning CNN-Transformer Model Architecture:

Table 3: Detailed architecture of CNN-Transformer deep learning model as provided by the *tensorflow.keras.Model.summary()* function. Total number of parameters is also displayed.

| Layer (type) | Output Shape | Param # | Connected to |
|---|---|---|---|
| input_1 (InputLayer) | [(None, 1280, 23, 1)] | 0 | [] |
| conv2d_1 (Conv2D) | (None, 1278, 21, 32) | 320 | [input_1[0][0]] |
| dropout_1 (Dropout) | (None, 1278, 21, 32) | 0 | [conv2d_1[0][0]] |
| batch_normalization_1 (BatchNorm) | (None, 1278, 21, 32) | 128 | [dropout_1[0][0]] |
| max_pooling2d_1 (MaxPooling2D) | (None, 639, 10, 32) | 0 | [batch_normalization_1[0][0]] |
| conv2d_2 (Conv2D) | (None, 637, 8, 64) | 18496 | [max_pooling2d_1[0][0]] |
| dropout_2 (Dropout) | (None, 637, 8, 64) | 0 | [conv2d_2[0][0]] |
| batch_normalization_2 (BatchNorm) | (None, 637, 8, 64) | 256 | [dropout_2[0][0]] |
| max_pooling2d_2 (MaxPooling2D) | (None, 318, 4, 64) | 0 | [batch_normalization_2[0][0]] |
| conv2d_3 (Conv2D) | (None, 316, 2, 128) | 73856 | [max_pooling2d_2[0][0]] |
| reshape_1 (Reshape) | (None, 316, 256) | 0 | [conv2d_3[0][0]] |
| multi_head_attention_1 (MultiHeadAttention) | (None, 316, 256) | 526080 | [reshape_1[0][0], reshape_1[0][0], reshape_1[0][0]] |
| layer_normalization_1 (LayerNorm) | (None, 316, 256) | 512 | [multi_head_attention_1[0][0]] |
| dropout_3 (Dropout) | (None, 316, 256) | 0 | [layer_normalization_1[0][0]] |
| dense_1 (Dense) | (None, 316, 64) | 16448 | [dropout_3[0][0]] |
| multi_head_attention_2 (MultiHeadAttention) | (None, 316, 64) | 132672 | [dense_1[0][0], dense_1[0][0], dense_1[0][0]] |
| layer_normalization_2 (LayerNorm) | (None, 316, 64) | 128 | [multi_head_attention_2[0][0]] |
| dropout_4 (Dropout) | (None, 316, 64) | 0 | [layer_normalization_2[0][0]] |
| dense_2 (Dense) | (None, 316, 64) | 4160 | [dropout_4[0][0]] |
| flatten_1 (Flatten) | (None, 20224) | 0 | [dense_2[0][0]] |
| dense_3 (Dense) | (None, 1) | 20225 | [flatten_1[0][0]] |

**Total params**: 793281
**Trainable params**: 793089
**Non-trainable params**: 192

# 3. Continuous Input-Output Performance Ratio:

**Intuition**: Observe the behavior of a seizure prediction model, subject to continuous unlabeled EEG prior to seizure onset to compare the effect of different data labelling criteria (i.e., different definitions for the preictal period).

**Input**: Output predictions of the model (0 or 1) for continuous multi-hour EEG, subject to an 8-minute non-overlapping averaging window (mean prediction value every 8 minutes).

**Output**: Quantitative measures of model's behavior including seizure prediction time, accuracy of predictions, transition period between interictal and preictal states, and output fluctuations. Computation of an integrative "performance score" for each preictal period definition.

**Method**: Fit a 4-parameter logistic sigmoidal curve to quantitatively model the classifier's output behavior. Key metrics, fitted curve, and smoothed output predictions displayed in Figure 2.

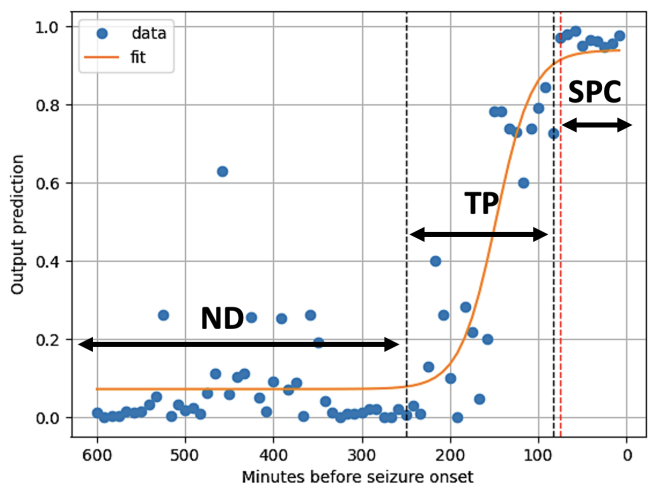

Figure 2: Continuous output predictions based on 600 minutes of EEG data preceding seizure onset, along with the fitted sigmoidal curve. The black dotted vertical lines, derived from the sigmoid fit, denote the start and end points of the transition period between interictal and preictal states. The red dotted line, directly calculated from the output, designates the start of the seizure prediction convergence, where the classifier's output converges. Notations: $ND$ = Negative Duration (interictal predictions), $TP$ = Transition Period, and $SPC$ = Seizure Prediction Convergence.

**Seizure Prediction Convergence (SPC)**: The point in time relative to seizure onset that the model's output converges. It is measured in minutes and represents the prediction horizon or prediction time in a seizure prediction system.

---

**Algorithm 1** Seizure Prediction Convergence ($SPC$)

1: **Input:** Prediction outputs over time
2: **Output:** $SPC$ time
3: **procedure** CALCULATESPC($predictions$)
4:     $meanPredictions \leftarrow$ MEAN($predictions$, 3)     ▷ Average over every 3 consecutive predictions
5:     $maxValue \leftarrow \max(meanPredictions)$
6:     $threshold \leftarrow 0.99 \times maxValue$
7:     **for** $i \leftarrow 1$ to length($meanPredictions$) **do**
8:         **if** $meanPredictions[i] \geq threshold$ **then**
9:             $SPC \leftarrow$ time($meanPredictions[i]$)     ▷ Time at 99% of max
10:            **return** $SPC$
11:        **end if**
12:    **end for**
13: **end procedure**

---

5

**Transition Period (TP)**: Duration of "transition period" between interictal and preictal states, measured in minutes. Together with the **Inflection Point (Infl)** of the fitted sigmoidal curve, it indicates "when" and "for how long" this transition takes place. *Infl* is measured in minutes before the seizure onset.

---

**Algorithm 2** Calculate Transition Period from Sigmoid Curve Amplitudes
---
1: **Input:** Sigmoid curve amplitudes $A$
2: **Output:** Transition period $TP$
3: **procedure** CALCULATETP($A$)
4:     $P_{low} \leftarrow \text{percentile}(A, 5)$                                                                  ▷ 5th percentile
5:     $P_{high} \leftarrow \text{percentile}(A, 95)$                                                                 ▷ 95th percentile
6:     $t_{low} \leftarrow \text{FIND\_INDEX}(A, P_{low})$
7:     $t_{high} \leftarrow \text{FIND\_INDEX}(A, P_{high})$
8:     $TP \leftarrow t_{high} - t_{low}$                                                       ▷ Calculate transition period
9:     **return** $TP$
10: **end procedure**
11: **function** FIND\_INDEX($A, P$)
12:     **for** $i \leftarrow 1$ **to** length($A$) **do**
13:         **if** $A[i] \geq P$ **then**
14:             **return** $i$
15:         **end if**
16:     **end for**
17: **end function**

---

**Negative Duration (ND)**: The duration (in minutes) that interictal predictions take place, defined as the period between the start of the output stream and the beginning of the $TP$. It is given by the equation below, where $D$ is the duration of the complete input EEG signal in minutes:

$$ND = D - \frac{TP}{2} - Infl. \qquad (1)$$

**SPC error ($SPC_{err}$) and ND error ($ND_{err}$)**: The average prediction error (deviation from "ideal" behavior) during the $SPC$ (deviation from 1) and $ND$ (deviation from 0) regions, respectively:

$$SPC_{err} = \frac{1}{N_{SPC}} \sum_{i=1}^{N_{SPC}} |1 - y_i| \qquad (2)$$

$$ND_{err} = \frac{1}{N_{ND}} \sum_{i=1}^{N_{ND}} |y_i| \qquad (3)$$

**Effective $SPC$ ($SPC_{eff}$) and $ND$ ($ND_{eff}$)**: The "effective" values for $SPC$ and $ND$, compensating for the average error during these regions, given by:

$$SPC_{eff} = SPC(1 - SPC_{err}) \qquad (4)$$

$$ND_{eff} = ND(1 - ND_{err}) \qquad (5)$$

**Continuous Input-Output Performance Coefficient (CIOPC)**: Metric that combines the aforementioned measures into a single value, by assigning weights based on the preferred usability. In this case, the metric is tuned to prioritize early prediction times at high accuracy, while having sharp transition between the two states. It is calculated by the equation below, where $\eta$ is the scaling coefficient that ensures $SPC_{eff}$ has at least the same weight with the second term, irrespective of the input signal duration. $ND_{eff}$ ensures that a short $TP$ is positively rewarded ($ND$ decreases with increasing $TP$) and that positive predictions farther from the onset are minimized.

$$CIOPC = SPC_{eff} + \eta(ND_{eff} + Infl_{comp}) \qquad (6)$$

The **scaling coefficient, $\eta$,** is consistent across all the different preictal period lengths within a given seizure, defined by the equation below. $SPC_{60}$, $SPC_{45}$, $SPC_{30}$, and $SPC_{15}$, refer to the $SPC$ of the model trained with a 60-, 45-, 30-, or a 15-minute preictal length respectively. The same holds for the $ND$ terms.

$$\eta = \begin{cases} 1, & \text{if } ND_{\max} < SPC_{\max} \\ \frac{SPC_{\max}}{ND_{\max}}, & \text{if } ND_{\max} \geq SPC_{\max} \end{cases}$$

where

$$SPC_{\max} = \max\{SPC_{60}, SPC_{45}, SPC_{30}, SPC_{15}\}$$

and

$$ND_{\max} = \max\{ND_{60}, ND_{45}, ND_{30}, ND_{15}\}$$

$ND$ also decreases with increasing $Infl$, which is undesired since $Infl$ is usually associated with earlier predictions and should be neither penalized nor rewarded (early predictions already rewarded by $SPC$). The introduction of the **Inflection Compensation** $Infl_{comp}$ term compensates for this, and is equivalent to the "loss" in $ND$ due to earlier point of inflection. Since $ND$ is scaled down by the average error to become $ND_{eff}$, $Infl_{comp}$ is also scaled down by the weighted-average of $SPC_{err}$ and $ND_{err}$. The detailed equation for computing the $Infl_{comp}$ is the following:

$$Inlf_{comp} = (Infl - Infl_{min}) \times Ave_{error} \quad \text{with} \quad Infl_{min} = \min\{Infl_{60}, Infl_{45}, Infl_{30}, Infl_{15}\}$$

$$\text{and} \quad Ave_{error} = \frac{SPC(1 - SPC_{err}) + ND(1 - ND_{err})}{SPC + ND} = \frac{SPC_{eff} + ND_{eff}}{SPC + ND}$$

**Continuous Input-Output Performance Ratio (CIOPR)**: Normalized form of the $CIOPC$ metric across the different preictal period lengths (e.g., 60, 45, 30, 15) to enable direct comparison. It also allows averaging the performance across different seizures, irrespective of the input duration, $D$, and $SPC$ that cause the $CIOPC$ to significantly vary among seizures and patients.

$$CIOPR_k = \frac{CIOPC_k}{CIOPC_{max}}, k \in \{60, 45, 30, 15\} \quad (7)$$

**Pearson Correlation ($\rho$)**: The correlation coefficient between the smoothed output predictions and the fitted sigmoidal curve. It can be used to assess the applicability of the proposed methodology, as well as to indirectly measure the extent of the output fluctuations (stable output leads to higher $\rho$ values).

# 4. Analytical CIOPR results:

Table 4: Analytical values of all CIOPR-related measures for each applicable seizure, including $SPC$, $SPC_{err}$, $Neg_{Err}$ (this page), and $Infl$, $TP$, $CIOPR$ (following page). The $\rho$ values are also displayed.

| Case | Seiz. ID | Time | SPC (mins) | | | | SPC Error | | | | Neg Error | | | |
|---|---|---|---|---|---|---|---|---|---|---|---|---|---|---|
| | | | 60 mins | 45 mins | 30 mins | 15 mins | 60 mins | 45 mins | 30 mins | 15 mins | 60 mins | 45 mins | 30 mins | 15 mins |
| chb01 | 1 | 14:33 | 72.2 | 72.9 | 72.9 | 75.0 | 0.0012 | 0.0014 | 0.0016 | 0.0041 | 0.0516 | 0.0295 | 0.0159 | 0.0027 |
| chb01 | 2 | 02:14 | 66.7 | 62.5 | 64.6 | 50.0 | 0.0004 | 0.0003 | 0.0020 | 0.0075 | 0.0423 | 0.0420 | 0.0617 | 0.0129 |
| chb01 | 5 | 13:05 | 54.2 | 50.0 | 33.3 | 33.3 | 0.0042 | 0.0007 | 0.0081 | 0.0758 | 0.7069 | 0.5663 | 0.2786 | 0.0427 |
| chb01 | Avg | | 64.3 | 61.8 | 56.9 | 52.8 | 0.0019 | 0.0008 | 0.0039 | 0.0291 | 0.2670 | 0.2126 | 0.1187 | 0.0194 |
| chb02 | 1 | 09:34 | 66.7 | 52.1 | 41.7 | 33.3 | 0.0112 | 0.0157 | 0.0129 | 0.1497 | 0.0227 | 0.0230 | 0.0179 | 0.0243 |
| chb02 | Avg | | 66.7 | 52.1 | 41.7 | 33.3 | 0.0112 | 0.0157 | 0.0129 | 0.1497 | 0.0227 | 0.0230 | 0.0179 | 0.0243 |
| chb04 | 1 | 12:12 | 133.3 | 133.3 | 133.3 | 133.3 | 0.0006 | 0.0022 | 0.0017 | 0.0045 | 0.0271 | 0.0338 | 0.0290 | 0.0435 |
| chb04 | 2 | 20:41 | 208.3 | 217.6 | 208.3 | 166.7 | 0.0032 | 0.0038 | 0.0053 | 0.0025 | - | - | - | - |
| chb04 | Avg | | 170.8 | 175.5 | 170.8 | 150.0 | 0.0019 | 0.0030 | 0.0035 | 0.0035 | 0.0250 | 0.0631 | 0.0298 | 0.0592 |
| chb05 | 1 | 22:28 | 62.5 | 58.3 | 41.7 | 33.3 | 0.0619 | 0.1112 | 0.0944 | 0.2652 | 0.6239 | 0.6647 | 0.6351 | 0.4648 |
| chb05 | 4 | 15:02 | 47.2 | 41.7 | 36.1 | 25.0 | 0.0674 | 0.0668 | 0.0616 | 0.0621 | 0.7372 | 0.6625 | 0.6340 | 0.6943 |
| chb05 | Avg | | 54.9 | 50.0 | 38.9 | 29.2 | 0.0646 | 0.0890 | 0.0780 | 0.1637 | 0.6805 | 0.6636 | 0.6346 | 0.5795 |
| chb06 | 4 | 07:15 | 36.1 | 36.1 | 25.0 | 16.7 | 0.1103 | 0.0961 | 0.0756 | 0.1142 | 0.2528 | 0.2622 | 0.1854 | 0.1585 |
| chb06 | 6 | 06:20 | 52.8 | 47.2 | 33.3 | 25.0 | 0.0942 | 0.0791 | 0.0645 | 0.1201 | 0.2108 | 0.1940 | 0.1960 | 0.1881 |
| chb06 | 7 | 09:52 | 58.3 | 33.3 | 33.3 | 16.7 | 0.1088 | 0.1180 | 0.0679 | 0.0929 | 0.7195 | 0.7056 | 0.6904 | 0.6233 |
| chb06 | 8 | 17:01 | 66.7 | 41.7 | 30.6 | 25.0 | 0.0466 | 0.0444 | 0.0400 | 0.0646 | 0.1127 | 0.1144 | 0.1425 | 0.1045 |
| chb06 | 9 | 13:56 | 53.5 | 39.6 | 30.6 | 20.8 | 0.0900 | 0.0844 | 0.0620 | 0.0980 | 0.3240 | 0.3190 | 0.3036 | 0.2686 |
| chb06 | 10 | 11:00 | 50.0 | 41.7 | 33.3 | 16.7 | 0.0246 | 0.0525 | 0.0150 | 0.0138 | 0.4315 | 0.2228 | 0.2290 | 0.0958 |
| chb06 | Avg | | 52.9 | 39.6 | 30.6 | 20.8 | 0.0900 | 0.0844 | 0.0620 | 0.0980 | 0.3240 | 0.3190 | 0.3036 | 0.2686 |
| chb07 | 1 | 09:08 | 57.5 | 50.0 | 43.3 | 36.7 | 0.0152 | 0.0128 | 0.0184 | 0.0359 | 0.0652 | 0.0301 | 0.0331 | 0.0494 |
| chb07 | 2 | 11:42 | 66.7 | 50.0 | 41.7 | 33.3 | 0.0221 | 0.0069 | 0.0450 | 0.0338 | 0.2639 | 0.2706 | 0.3003 | 0.2500 |
| chb07 | 3 | 12:00 | 60.4 | 58.3 | 33.3 | 25.0 | 0.0099 | 0.0312 | 0.0121 | 0.0269 | 0.0492 | 0.0204 | 0.0090 | 0.0526 |
| chb07 | Avg | | 61.5 | 52.8 | 39.4 | 31.7 | 0.0157 | 0.0170 | 0.0252 | 0.0322 | 0.1261 | 0.1070 | 0.1141 | 0.1173 |
| chb09 | 1 | 15:45 | 125.0 | 97.2 | 66.7 | 41.7 | 0.0037 | 0.0061 | 0.0059 | 0.0084 | 0.0651 | 0.0700 | 0.1049 | 0.0613 |
| chb09 | 2 | 21:11 | 91.7 | 83.3 | 83.3 | 83.3 | 0.0074 | 0.0029 | 0.0155 | 0.0349 | 0.0645 | 0.0556 | 0.2475 | 0.1831 |
| chb09 | 4 | 14:03 | 63.9 | 63.9 | 41.7 | 25.0 | 0.0161 | 0.0182 | 0.0283 | 0.0333 | 0.0258 | 0.0244 | 0.0314 | 0.0369 |
| chb09 | Avg | | 93.5 | 81.5 | 63.9 | 50.0 | 0.0091 | 0.0091 | 0.0166 | 0.0255 | 0.0518 | 0.0500 | 0.1280 | 0.0938 |
| chb10 | 1 | 09:01 | 70.8 | 62.5 | 43.8 | 25.0 | 0.0363 | 0.0452 | 0.0599 | 0.1630 | 0.0795 | 0.0792 | 0.0502 | 0.0441 |
| chb10 | Avg | | 70.8 | 62.5 | 43.8 | 25.0 | 0.0363 | 0.0452 | 0.0599 | 0.1630 | 0.0795 | 0.0792 | 0.0502 | 0.0441 |
| chb14 | 1 | 17:08 | 50.0 | 41.7 | 33.3 | 16.7 | 0.0373 | 0.0936 | 0.0126 | 0.0354 | 0.6983 | 0.7279 | 0.7137 | 0.6766 |
| chb14 | 5 | 17:24 | 54.2 | 50.0 | 33.3 | 25.0 | 0.0260 | 0.0706 | 0.0207 | 0.0791 | 0.4120 | 0.4076 | 0.4082 | 0.3665 |
| chb14 | Avg | | 52.1 | 45.8 | 33.3 | 20.8 | 0.0317 | 0.0821 | 0.0166 | 0.0572 | 0.5551 | 0.5678 | 0.5609 | 0.5216 |
| chb16 | 1 | 02:18 | 62.5 | 43.8 | 33.3 | 41.7 | 0.0327 | 0.0568 | 0.0321 | 0.1707 | 0.0491 | 0.0557 | 0.1027 | 0.0821 |
| chb16 | Avg | | 62.5 | 43.8 | 33.3 | 41.7 | 0.0327 | 0.0568 | 0.0321 | 0.1707 | 0.0491 | 0.0557 | 0.1027 | 0.0821 |
| chb18 | 1 | 06:57 | 56.3 | 50.0 | 33.3 | 33.3 | 0.0148 | 0.0127 | 0.0184 | 0.1080 | 0.0645 | 0.0962 | 0.0740 | 0.0956 |
| chb18 | Avg | | 56.3 | 50.0 | 33.3 | 33.3 | 0.0148 | 0.0127 | 0.0184 | 0.1080 | 0.0645 | 0.0962 | 0.0740 | 0.0956 |
| chb21 | 1 | 12:20 | 139.6 | 139.6 | 125.0 | 139.6 | 0.0303 | 0.0303 | 0.0358 | 0.0753 | 0.0081 | 0.0081 | 0.0089 | 0.0102 |
| chb21 | Avg | | 139.6 | 139.6 | 125.0 | 139.6 | 0.0303 | 0.0303 | 0.0358 | 0.0753 | 0.0081 | 0.0081 | 0.0089 | 0.0102 |
| chb22 | 1 | 18:41 | 66.7 | 58.3 | 33.3 | 25.0 | 0.1463 | 0.2640 | 0.3201 | 0.3488 | 0.1608 | 0.1988 | 0.1102 | 0.2256 |
| chb22 | 2 | 23:37 | 50.0 | 41.7 | 25.0 | 25.0 | 0.0126 | 0.0068 | 0.0013 | 0.0788 | 0.0010 | 0.0054 | 0.0003 | 0.1275 |
| chb22 | Avg | | 58.3 | 50.0 | 29.2 | 25.0 | 0.0795 | 0.1354 | 0.1607 | 0.2138 | 0.0809 | 0.1021 | 0.0552 | 0.1765 |
| Mean | - | | 73.0 | 64.9 | 53.2 | 44.5 | 0.0370 | 0.0477 | 0.0407 | 0.0824 | 0.2322 | 0.2181 | 0.2079 | 0.1847 |
| ± Std | | | 35.8 | 38.9 | 40.0 | 38.8 | 0.0389 | 0.0555 | 0.0597 | 0.0794 | 0.2504 | 0.2378 | 0.2200 | 0.2025 |

| Case | Inflection Point (mins) | | | | Transition Period (mins) | | | | Pearson Correlation | | | | CIOPR | | | |
|---|---|---|---|---|---|---|---|---|---|---|---|---|---|---|---|---|
| | 60 mins | 45 mins | 30 mins | 15 mins | 60 mins | 45 mins | 30 mins | 15 mins | 60 mins | 45 mins | 30 mins | 15 mins | 60 mins | 45 mins | 30 mins | 15 mins |
| chb01 | 83.7 | 83.7 | 83.9 | 83.8 | 16.7 | 16.7 | 16.7 | 16.7 | 0.9885 | 0.9959 | 0.9990 | 1.0000 | 0.9582 | 0.9739 | 0.9806 | 1.0000 |
| chb01 | 86.0 | 81.8 | 81.3 | 69.6 | 54.2 | 29.2 | 33.3 | 29.2 | 0.9832 | 0.9833 | 0.9834 | 0.9980 | 1.0000 | 0.9803 | 0.9836 | 0.8954 |
| chb01 | 70.8 | 67.3 | 59.5 | 49.4 | 20.8 | 16.7 | 45.8 | 58.3 | 0.9186 | 0.9156 | 0.9761 | 0.9895 | 0.9763 | 1.0000 | 0.8348 | 0.8558 |
| chb01 | 80.2 | 77.6 | 74.9 | 67.6 | 30.6 | 20.8 | 31.9 | 34.7 | 0.9634 | 0.9649 | 0.9862 | 0.9958 | 0.9782 | 0.9847 | 0.9330 | 0.9171 |
| chb02 | 74.0 | 72.6 | 72.4 | 50.2 | 16.7 | 16.7 | 16.7 | 81.3 | 0.9941 | 0.9952 | 0.9961 | 0.9828 | 1.0000 | 0.8924 | 0.8211 | 0.6644 |
| chb02 | 74.0 | 72.6 | 72.4 | 50.2 | 16.7 | 16.7 | 16.7 | 81.3 | 0.9941 | 0.9952 | 0.9961 | 0.9828 | 1.0000 | 0.8924 | 0.8211 | 0.6644 |
| chb04 | 177.5 | 173.2 | 173.6 | 208.5 | 218.8 | 187.5 | 208.3 | 225.0 | 0.9512 | 0.9531 | 0.9514 | 0.9521 | 0.9810 | 1.0000 | 0.9873 | 0.9676 |
| chb04 | 208.3 | 217.6 | 217.3 | 180.5 | 14.2 | 12.5 | 12.5 | 12.5 | 0.9611 | 0.9686 | 0.9426 | 0.6161 | 0.9578 | 1.0000 | 0.9558 | 0.7671 |
| chb04 | 192.9 | 195.4 | 195.5 | 194.5 | 116.5 | 100.0 | 110.4 | 118.8 | 0.9561 | 0.9609 | 0.9470 | 0.7841 | 0.9694 | 1.0000 | 0.9716 | 0.8674 |
| chb05 | 151.3 | 154.3 | 151.0 | 141.6 | 33.3 | 16.7 | 25.0 | 16.7 | 0.7128 | 0.7254 | 0.7084 | 0.6051 | 1.0000 | 0.9038 | 0.7412 | 0.7059 |
| chb05 | 117.8 | 111.6 | 91.3 | 26.1 | 88.9 | 130.6 | 158.3 | 16.7 | 0.7491 | 0.7455 | 0.6543 | 0.5942 | 1.0000 | 0.9150 | 0.7881 | 0.5814 |
| chb05 | 134.5 | 132.9 | 121.2 | 83.8 | 61.1 | 73.6 | 91.7 | 16.7 | 0.7309 | 0.7354 | 0.6814 | 0.5996 | 1.0000 | 0.9094 | 0.7647 | 0.6437 |
| chb06 | 65.6 | 49.7 | 41.2 | 23.8 | 33.3 | 22.2 | 16.7 | 36.1 | 0.8640 | 0.8352 | 0.8134 | 0.7229 | 0.9912 | 1.0000 | 0.8903 | 0.7516 |
| chb06 | 87.2 | 72.0 | 61.3 | 52.7 | 63.9 | 88.9 | 102.8 | 91.7 | 0.9102 | 0.8849 | 0.8374 | 0.8043 | 1.0000 | 0.9492 | 0.8054 | 0.7139 |
| chb06 | 66.5 | 50.1 | 34.4 | 25.3 | 16.7 | 16.7 | 8.3 | 19.4 | 0.7938 | 0.7011 | 0.7113 | 0.8205 | 1.0000 | 0.6542 | 0.6705 | 0.4879 |
| chb06 | 266.1 | 266.0 | 268.2 | 273.0 | 69.4 | 54.2 | 55.6 | 22.2 | 0.9894 | 0.9895 | 0.9779 | 0.9872 | 1.0000 | 0.6257 | 0.4944 | 0.4700 |
| chb06 | 121.3 | 109.4 | 101.3 | 93.7 | 45.8 | 45.5 | 45.8 | 42.4 | 0.8893 | 0.8527 | 0.8350 | 0.8337 | 1.0000 | 0.8539 | 0.7727 | 0.6791 |
| chb06 | 61.0 | 50.6 | 50.7 | 18.3 | 25.0 | 16.7 | 16.7 | 66.7 | 0.8949 | 0.9649 | 0.9564 | 0.9108 | 0.9842 | 1.0000 | 0.9218 | 0.6936 |
| chb06 | 121.3 | 109.4 | 101.3 | 93.7 | 45.8 | 45.5 | 45.8 | 42.4 | 0.8893 | 0.8527 | 0.8350 | 0.8337 | 0.9959 | 0.8472 | 0.7592 | 0.6327 |
| chb07 | 67.6 | 66.8 | 67.1 | 67.2 | 16.7 | 20.0 | 20.0 | 16.7 | 0.9473 | 0.9782 | 0.9763 | 0.9568 | 1.0000 | 0.9508 | 0.8871 | 0.8158 |
| chb07 | 74.9 | 74.4 | 74.0 | 74.8 | 16.7 | 16.7 | 16.7 | 16.7 | 0.9706 | 0.9677 | 0.9450 | 0.9673 | 1.0000 | 0.8598 | 0.7559 | 0.7193 |
| chb07 | 237.2 | 228.3 | 216.6 | 237.9 | 177.1 | 218.8 | 208.3 | 170.8 | 0.9660 | 0.9625 | 0.9444 | 0.9575 | 1.0000 | 0.8893 | 0.7332 | 0.7307 |
| chb07 | 126.5 | 123.2 | 119.3 | 126.6 | 70.1 | 85.1 | 81.7 | 68.1 | 0.9613 | 0.9695 | 0.9552 | 0.9606 | 1.0000 | 0.9000 | 0.7921 | 0.7553 |
| chb09 | 192.1 | 191.8 | 191.8 | 167.3 | 16.7 | 16.7 | 16.7 | 180.6 | 0.9499 | 0.9662 | 0.8830 | 0.8892 | 1.0000 | 0.8850 | 0.7448 | 0.5701 |
| chb09 | 214.0 | 215.3 | 216.2 | 214.6 | 25.0 | 16.7 | 16.7 | 25.0 | 0.9966 | 0.9981 | 0.9860 | 0.9789 | 1.0000 | 0.9715 | 0.9022 | 0.8814 |
| chb09 | 335.0 | 335.4 | 329.4 | 334.7 | 19.4 | 16.7 | 186.1 | 16.7 | 0.9893 | 0.9880 | 0.9642 | 0.9778 | 0.9942 | 1.0000 | 0.4405 | 0.6925 |
| chb09 | 247.1 | 247.5 | 245.8 | 238.8 | 20.4 | 16.7 | 73.2 | 74.1 | 0.9786 | 0.9841 | 0.9444 | 0.9486 | 0.9981 | 0.9522 | 0.6958 | 0.7147 |
| chb10 | 144.2 | 145.6 | 127.7 | 111.9 | 143.8 | 100.0 | 162.5 | 204.2 | 0.9326 | 0.9414 | 0.9493 | 0.9276 | 1.0000 | 0.9628 | 0.8035 | 0.6315 |
| chb10 | 144.2 | 145.6 | 127.7 | 111.9 | 143.8 | 100.0 | 162.5 | 204.2 | 0.9326 | 0.9414 | 0.9493 | 0.9276 | 1.0000 | 0.9628 | 0.8035 | 0.6315 |
| chb14 | 73.6 | 50.6 | 41.5 | 24.4 | 33.3 | 16.7 | 16.7 | 16.7 | 0.9121 | 0.8437 | 0.9068 | 0.8399 | 1.0000 | 0.7731 | 0.6955 | 0.4538 |
| chb14 | 67.2 | 58.1 | 50.8 | 35.5 | 16.7 | 16.7 | 39.6 | 58.3 | 0.8720 | 0.8554 | 0.8338 | 0.8238 | 1.0000 | 0.9240 | 0.7424 | 0.6345 |
| chb14 | 70.4 | 54.3 | 46.2 | 30.0 | 25.0 | 16.7 | 28.1 | 37.5 | 0.8920 | 0.8496 | 0.8703 | 0.8318 | 1.0000 | 0.8486 | 0.7190 | 0.5442 |
| chb16 | 170.3 | 99.3 | 117.0 | 115.6 | 220.8 | 247.9 | 247.9 | 289.6 | 0.9635 | 0.8992 | 0.9359 | 0.9046 | 1.0000 | 0.8215 | 0.7237 | 0.7261 |
| chb16 | 170.3 | 99.3 | 117.0 | 115.6 | 220.8 | 247.9 | 247.9 | 289.6 | 0.9635 | 0.8992 | 0.9359 | 0.9046 | 1.0000 | 0.8215 | 0.7237 | 0.7261 |
| chb18 | 77.3 | 77.7 | 77.0 | 77.0 | 25.0 | 27.1 | 25.0 | 30.6 | 0.9442 | 0.9180 | 0.9382 | 0.8814 | 1.0000 | 0.9257 | 0.7846 | 0.7432 |
| chb18 | 77.3 | 77.7 | 77.0 | 77.0 | 25.0 | 27.1 | 25.0 | 30.6 | 0.9442 | 0.9180 | 0.9382 | 0.8814 | 1.0000 | 0.9257 | 0.7846 | 0.7432 |
| chb21 | 253.5 | 253.5 | 244.2 | 240.9 | 133.3 | 133.3 | 110.4 | 106.3 | 0.9959 | 0.9959 | 0.9955 | 0.9935 | 1.0000 | 1.0000 | 0.9637 | 0.9985 |
| chb21 | 253.5 | 253.5 | 244.2 | 240.9 | 133.3 | 133.3 | 110.4 | 106.3 | 0.9959 | 0.9959 | 0.9955 | 0.9935 | 1.0000 | 1.0000 | 0.9637 | 0.9985 |
| chb22 | 282.3 | 291.1 | 293.7 | 301.4 | 100.0 | 16.7 | 25.0 | 50.0 | 0.8764 | 0.8338 | 0.7961 | 0.6869 | 0.8477 | 1.0000 | 0.8121 | 0.5709 |
| chb22 | 180.7 | 179.9 | 192.4 | 207.9 | 25.0 | 33.3 | 158.3 | 200.0 | 0.9662 | 0.9697 | 0.9119 | 0.9179 | 1.0000 | 0.8600 | 0.4047 | 0.5277 |
| chb22 | 231.5 | 235.5 | 243.0 | 254.7 | 62.5 | 25.0 | 91.7 | 125.0 | 0.9213 | 0.9018 | 0.8540 | 0.8024 | 0.9239 | 0.9300 | 0.6084 | 0.5493 |
| Mean | 143.1 | 136.7 | 133.1 | 125.3 | 60.3 | 56.0 | 71.9 | 75.6 | 0.9244 | 0.9153 | 0.9039 | 0.8757 | 0.9890 | 0.9133 | 0.7872 | 0.7118 |
| ± Std | 78.5 | 82.7 | 84.2 | 92.6 | 61.4 | 66.0 | 73.7 | 77.1 | 0.0722 | 0.0848 | 0.0935 | 0.1238 | 0.0296 | 0.0975 | 0.1489 | 0.1480 |

## 5. Statistical Analysis Results:

Table 5: Statistical comparisons of different preictal period lengths (minutes) using the non-parametric . Each row tests the null hypothesis that the two sample distributions are the same. P-values are displayed, after being adjusted by the Bonferroni correction for multiple tests. The significance level is 0.05. Each row represents a different metric for which the test is performed.

| Comparison | SPC | SPCerr | NDerr | Point of Infl. | TP | Pearson Corr. | CIOPR | Sensitivity | Specificity | Accuracy | F1-score |
|---|---|---|---|---|---|---|---|---|---|---|---|
| Fifteen-Thirty | 0.420 | 1.000 | 1.000 | 1.000 | 1.000 | 1.000 | 0.295 | 1.000 | 0.539 | 0.357 | 0.790 |
| Fifteen-Fortyfive | 0.000 | 1.000 | 1.000 | 0.178 | 0.652 | 0.005 | 0.000 | 0.891 | 0.034 | 0.041 | 0.266 |
| Fifteen-Sixty | 0.000 | 0.000 | 0.614 | 0.000 | 0.261 | 0.005 | 0.000 | 0.196 | 0.001 | 0.001 | 0.003 |
| Thirty-Fortyfive | 0.031 | 1.000 | 1.000 | 1.000 | 1.000 | 0.156 | 0.003 | 1.000 | 1.000 | 1.000 | 1.000 |
| Thirty-Sixty | 0.000 | 0.001 | 1.000 | 0.023 | 1.000 | 0.156 | 0.000 | 1.000 | 0.196 | 0.411 | 0.266 |
| Fortyfive-Sixty | 0.156 | 0.005 | 1.000 | 0.420 | 1.000 | 1.000 | 1.000 | 1.000 | 1.000 | 1.000 | 0.790 |
| Sig. Comparisons | 4 | 3 | 0 | 2 | 0 | 2 | 4 | 0 | 2 | 2 | 1 |
| p-value | < 0.001 | < 0.001 | 0.388 | < 0.001 | 0.135 | < 0.001 | < 0.001 | 0.161 | < 0.001 | 0.002 | 0.005 |

# 6. Comparison with State-Of-The-Art Literature:

Table 6: Balanced Accuracy (%) comparison between state-of-the-art literature and the proposed solution. The table includes studies reporting both sensitivity and specificity measures for at least five cases overlapping with our work.

| Case | Tsiouris et al. [1] | Wang et al. [2] | Ryu et al. [3] | Jemal et al. [4] | Baghdadi et al. [5] | Affes et al. [6] | Ma et al. [7] | Qi et al. [8] | Yang et al. [9] | Rasheed et al. [10] | Jana et al. [11] | Ibrahim et al. [12] | Average | Ours |
|---|---|---|---|---|---|---|---|---|---|---|---|---|---|---|
| chb01 | 100.00 | 99.00 | 100.00 | 96.75 | 93.00 | 71.77 | 99.99 | 97.81 | 99.65 | 86.30 | 96.15 | 97.60 | 94.04 | 99.93 |
| chb02 | 100.00 | 99.00 | 86.94 | 95.55 | 96.00 | 70.82 | 89.48 | 96.12 | 75.75 | 87.10 | 89.04 | 99.22 | 88.58 | 98.04 |
| chb03 | 100.00 | 100.00 | 96.82 | 98.80 | 93.00 | 69.74 | 97.74 | 88.25 | 95.90 | 90.56 | 96.48 | 77.90 | 92.73 | 98.71 |
| chb04 | 99.97 | 99.27 | 78.26 | 91.35 | 96.00 | 68.57 | N/A | N/A | N/A | N/A | 92.59 | 78.86 | 87.67 | 99.52 |
| chb05 | 99.83 | 99.67 | 94.33 | 85.00 | 89.00 | 68.32 | 97.11 | 90.38 | 93.50 | 87.55 | 83.76 | N/A | 88.86 | 98.14 |
| chb06 | 99.34 | 99.00 | 94.20 | 84.75 | 74.00 | 68.63 | 88.89 | N/A | N/A | N/A | 84.65 | 63.07 | 84.87 | 84.15 |
| chb07 | 99.88 | 100.00 | 100.00 | 93.60 | 99.00 | 68.20 | 83.97 | N/A | N/A | N/A | 90.28 | N/A | 90.72 | 98.64 |
| chb09 | 100.00 | 99.85 | 99.83 | 87.75 | 95.00 | 69.65 | N/A | 91.80 | 79.05 | 82.12 | 87.75 | N/A | 88.09 | 99.28 |
| chb10 | 99.91 | 97.83 | 90.53 | 90.70 | 82.50 | 70.57 | 93.37 | 90.42 | 80.85 | 86.28 | 72.14 | N/A | 85.52 | 95.62 |
| chb11 | 100.00 | 99.94 | 100.00 | 99.30 | 92.50 | 69.77 | 99.65 | N/A | N/A | N/A | N/A | 59.35 | 93.52 | 99.83 |
| chb14 | 99.86 | 100.00 | 89.67 | 74.70 | 76.50 | 73.79 | 89.93 | 90.56 | 69.30 | 92.01 | N/A | 71.45 | 84.05 | 92.33 |
| chb16 | 99.75 | 99.42 | 81.03 | 89.75 | 83.50 | 78.33 | N/A | N/A | N/A | N/A | N/A | N/A | 86.40 | 98.32 |
| chb17 | 99.95 | 99.76 | 100.00 | 97.05 | 94.50 | 76.31 | N/A | N/A | N/A | N/A | N/A | N/A | 93.52 | 98.51 |
| chb18 | 99.66 | 99.87 | 92.35 | 96.10 | 95.50 | 78.64 | N/A | 94.33 | 96.10 | 87.21 | N/A | N/A | 92.51 | 98.00 |
| chb19 | 100.00 | 98.65 | 100.00 | 99.50 | 95.50 | 74.95 | 98.83 | 95.97 | 95.80 | 80.69 | N/A | N/A | 93.32 | 99.82 |
| chb20 | 100.00 | 98.89 | 100.00 | 98.90 | 89.00 | 77.30 | 99.95 | 97.01 | 99.35 | 84.79 | N/A | N/A | 93.91 | 99.39 |
| chb21 | 100.00 | 98.98 | 96.91 | 89.10 | 95.50 | 77.14 | 99.55 | 97.85 | 92.70 | 96.79 | N/A | N/A | 93.39 | 98.37 |
| chb22 | 100.00 | 97.36 | 91.46 | 88.35 | 95.50 | 73.54 | 93.19 | N/A | N/A | N/A | N/A | N/A | 89.90 | 92.79 |
| chb23 | 100.00 | 95.00 | 94.64 | 94.75 | 88.00 | 55.95 | 99.75 | 99.52 | 98.90 | 97.50 | 85.31 | N/A | 90.93 | 99.68 |
| Mean | 99.90 | 99.02 | 94.05 | 92.20 | 90.71 | 71.68 | 94.81 | 94.17 | 89.74 | 88.24 | 87.81 | 78.20 | 90.13 | 97.32 |
| ± Std | 0.16 | 1.16 | 6.22 | 6.07 | 6.62 | 4.99 | 4.77 | 3.43 | 9.68 | 4.77 | 6.42 | 13.46 | 3.17 | 3.76 |

Table 7: Prediction times (minutes) across the literature.

| **Work** | Khan et al. [13] | Zandi et al. [14] | Tang et al. [15] | Wang et al. [2] |
|---|---|---|---|---|
| Prediction time (mins) | 5.8 | 22.5 | 27.2 | 29.5 |
| **Work** | Zhang et al. [16] | Chu et al. [17] | Otaiby et al. [18] | **Ours** |
| Prediction time (mins) | 43.9 | 45.3 | 68.7 | 76.8 |



\end{document}